\documentclass[letterpaper,twocolumn,10pt]{article}
\usepackage{usenix2019_v3}

\usepackage{tikz}
\usepackage{filecontents}
\usepackage{xcolor}
\usepackage{subcaption}
\usepackage{amsmath,amssymb,amsfonts}
\DeclareMathOperator*{\argmax}{argmax}
\DeclareMathOperator*{\argmin}{argmin}
\DeclareMathOperator*{\mentropy}{Mentr}
\usepackage{dsfont}
\newcommand{\bfx}{\mathbf{x}}
\newcommand{\bfz}{\mathbf{z}}
\newcommand{\bfn}{\mathbf{n}}
\newcommand{\advobserv}{\mathcal{O}(F, \mathbf{z})}

\newtheorem{definition}{Definition}
\usepackage{booktabs} 
\usepackage{multirow}

\begin{document}

\date{}

\title{\Large \bf Systematic Evaluation of Privacy Risks of Machine Learning Models 
}

\author{
{\rm Liwei Song}\\
{\rm liweis@princeton.edu}\\
Princeton University
\and
{\rm Prateek Mittal}\\
{\rm pmittal@princeton.edu}\\
Princeton University
} 

\maketitle

\begin{abstract}
Machine learning models are prone to memorizing sensitive data, making them vulnerable to membership inference attacks in which an adversary aims to guess if an input sample was used to train the model. 
In this paper, we show that prior work on membership inference attacks may severely underestimate the privacy risks by relying solely on training custom neural network classifiers to perform attacks and focusing only on the aggregate results over data samples, such as the attack accuracy.
To overcome these limitations, we first propose to benchmark membership inference privacy risks by improving existing non-neural network based inference attacks and proposing a new inference attack method based on a modification of prediction entropy.
We propose to supplement existing neural network based attacks with our proposed benchmark attacks to effectively measure the privacy risks.
We also propose benchmarks for defense mechanisms by accounting for adaptive adversaries with knowledge of the defense and also accounting for the trade-off between model accuracy and privacy risks.
Using our benchmark attacks, we demonstrate that existing defense approaches against membership inference attacks are not as effective as previously reported.

Next, we introduce a new approach for fine-grained privacy analysis by formulating and deriving a new metric called the privacy risk score. Our privacy risk score metric measures an individual sample's likelihood of being a training member, which allows an adversary to identify samples with high privacy risks and perform membership inference attacks with high confidence.
We propose to combine both existing aggregate privacy analysis and our proposed fine-grained privacy analysis for systematically measuring privacy risks.
We experimentally validate the effectiveness of the privacy risk score metric and demonstrate that the distribution of privacy risk scores across individual samples is heterogeneous. 
Finally, we perform an in-depth investigation to understand
why certain samples have high privacy risk scores, including correlations with model properties such as model sensitivity, generalization error, and feature embeddings.
Our work emphasizes the importance of a systematic and rigorous evaluation of privacy risks of machine learning models.
We publicly release our code at \url{https://github.com/inspire-group/membership-inference-evaluation} and our evaluation mechanisms have also been integrated in Google's TensorFlow Privacy library.

\end{abstract}
\section{Introduction}\label{sec:intro}

A recent thread of research has shown that machine learning (ML) models memorize sensitive information of training data, indicating serious privacy risks \cite{shokri_membership_SP17, fredrikson_inversion_CCS15, song_active_privacy_CCS17, ganju_property_privacy_CCS18, hitaj2017deep, carlini_memorization_usenix19, salem2019updates}.
In this paper, we focus on the membership inference attack, where the adversary aims to guess whether an input sample was used to train the target machine learning model or not \cite{shokri_membership_SP17, yeom_privacy_CSF18}.
It poses a severe privacy risk as the membership can reveal an individual’s sensitive information \cite{pyrgelis_mem_location_ndss18, backes2016membership}. 
For example, participation in a hospital’s health analytic training set means that an individual was once a patient in that hospital.
As membership inference attacks expose the privacy risks of an individual user participating in the training data, they serve as a valuable tool to quantify the privacy provided by differential privacy implementations~\cite{jayaraman_evaluate_dp_usenix19} and to help to guide 
the selection of privacy parameters in the broader class of statistical privacy frameworks~\cite{liu2019investigating}.
Shokri et al. \cite{shokri_membership_SP17} 
conducted membership inference attacks against machine learning classifiers in the black-box manner, where the adversary only observes prediction outputs of the target model.
They formalize the attack as a classification problem and train dedicated neural network (NN) classifiers to distinguish between training members and non-members.
The research community has since extended the idea of membership inference attacks to generative models \cite{hayes_membership_PETS19, hilprecht_mem_gan_pets19, chen_mem_gan_priml19, wu2019generalization}, to differentially private models \cite{jayaraman_evaluate_dp_usenix19, rahman2018membership}, to decentralized settings where the models are trained across multiple users without sharing their data \cite{nasr_whitebox_privacy_SP19, melis_mem_decentralized_SP19}, and to white-box settings where the adversary also has the access to the target model's architecture and weights \cite{nasr_whitebox_privacy_SP19}.

To mitigate such privacy risks, several defenses against membership inference attacks have been proposed.
Nasr et al. \cite{nasr_membership_defense_CCS18} propose to include membership inference attacks during the training process: they train the target model to simultaneously achieve correct predictions and low membership inference attack accuracy by adding the inference attack as an adversarial regularization term.
Jia et al. \cite{jia2019memguard_ccs19} propose a defense method called MemGuard
which does not require retraining the model: the model prediction outputs are obfuscated with noisy perturbations such that the adversary cannot distinguish between members and non-members based on the perturbed outputs.
Both papers show that their defenses greatly mitigate membership inference privacy risks, resulting in attack performance that is close to random guessing.

In this paper, we critically examine how previous work \cite{shokri_membership_SP17, nasr_membership_defense_CCS18, jia2019memguard_ccs19, nasr_whitebox_privacy_SP19, salem_membership_NDSS19} has evaluated the membership inference privacy risks of machine learning models, and demonstrate two key limitations that lead to a severe underestimation of privacy risks.
\emph{First, many prior papers, particularly those proposing defense methods \cite{nasr_membership_defense_CCS18, jia2019memguard_ccs19}, solely rely on training custom NN classifiers to perform membership inference attacks.}
These NN attack classifiers may underestimate privacy risks due to inappropriate settings of hyperparameters such as number of hidden layers and learning rate.
\emph{Second, existing evaluations only focus on aggregate notions of privacy risks faced by all data samples, lacking a fine-grained understanding of privacy risks faced by individual samples.}

\begin{table}[!ht]
\caption{Benchmarking the effectiveness of existing defenses \cite{nasr_membership_defense_CCS18, jia2019memguard_ccs19} against membership inference attacks.
Both Nasr et al. \cite{nasr_membership_defense_CCS18} and Jia et al. \cite{jia2019memguard_ccs19} report that for their defended models, custom NN classifiers achieve attack accuracy close to $50\%$, which is the accuracy of random guessing.
By using a suite of non-NN based attacks as our benchmark,
we find that the attack accuracy is significantly larger than previous estimates, ranging from an increase of $7.6\%$ to $23.9\%$.
}
\centering
\renewcommand\arraystretch{1.0}
\fontsize{9pt}{9pt}\selectfont
\begin{tabular}{cccc}
\toprule[1.5pt]
\multirow{2}{*}{\textbf{defense methods}} & \multirow{2}{*}{\textbf{dataset}} & \textbf{reported} & \textbf{our benchmark} \\
& & \textbf{attack acc} & \textbf{attack acc}\\
\midrule[0.75pt]
& \multirow{2}{*}{Purchase100} & \multirow{2}{*}{51.6\%} & \multirow{2}{*}{59.5\%} \\
{adversarial}  & & & \\
{regularization \cite{nasr_membership_defense_CCS18}} & \multirow{2}{*}{Texas100} & \multirow{2}{*}{51.0\%} & \multirow{2}{*}{58.6\%}  \\
& & & \\
\midrule[0.75pt]
\multirow{4}{*}{MemGuard \cite{jia2019memguard_ccs19}} & \multirow{2}{*}{Location30} & \multirow{2}{*}{50.1\%} & \multirow{2}{*}{69.1\%} \\
& & & \\
 & \multirow{2}{*}{Texas100} & \multirow{2}{*}{50.3\%} & \multirow{2}{*}{74.2\%} \\
 & & & \\
\bottomrule[1.5pt]
\end{tabular}
\label{tab:benchmark_summary}
\end{table}

To overcome the limitation of reliance on NN-based attacks, we propose to use a suite of alternative existing and novel non-NN based attack methods to benchmark the membership inference privacy risks.
These benchmark attack methods make inference decisions based on computing custom metrics on the predictions of the target model.
Compared to NN-based attacks, our proposed benchmark attacks are easy to implement without hyperparameter tuning. We only need to set the threshold values using the shadow-training technique~\cite{shokri_membership_SP17}.
We also show that rigorously benchmarking defense mechanisms requires a careful consideration of strategic adversaries aware of the defense mechanism, as well as alternative baselines that trade-off accuracy of the target machine learning model with privacy risks.
With our proposed benchmark attacks, we indeed find that that existing membership inference defense methods \cite{nasr_membership_defense_CCS18, jia2019memguard_ccs19} are not as effective as previously reported.
As shown in Table~\ref{tab:benchmark_summary}, the adversary can still perform membership inference attacks on models defended by adversarial regularization \cite{nasr_membership_defense_CCS18} and MemGuard \cite{jia2019memguard_ccs19} with an accuracy ranging from $58.6\%$ to $74.2\%$, instead of the reported accuracy around $50\%$, which is the accuracy of random guessing. 
Therefore, we argue that these non-NN based attacks should \emph{supplement} existing NN based attacks to effectively measure the privacy risks.

\begin{figure}[!ht]
	\centering
	\includegraphics[width=\linewidth]{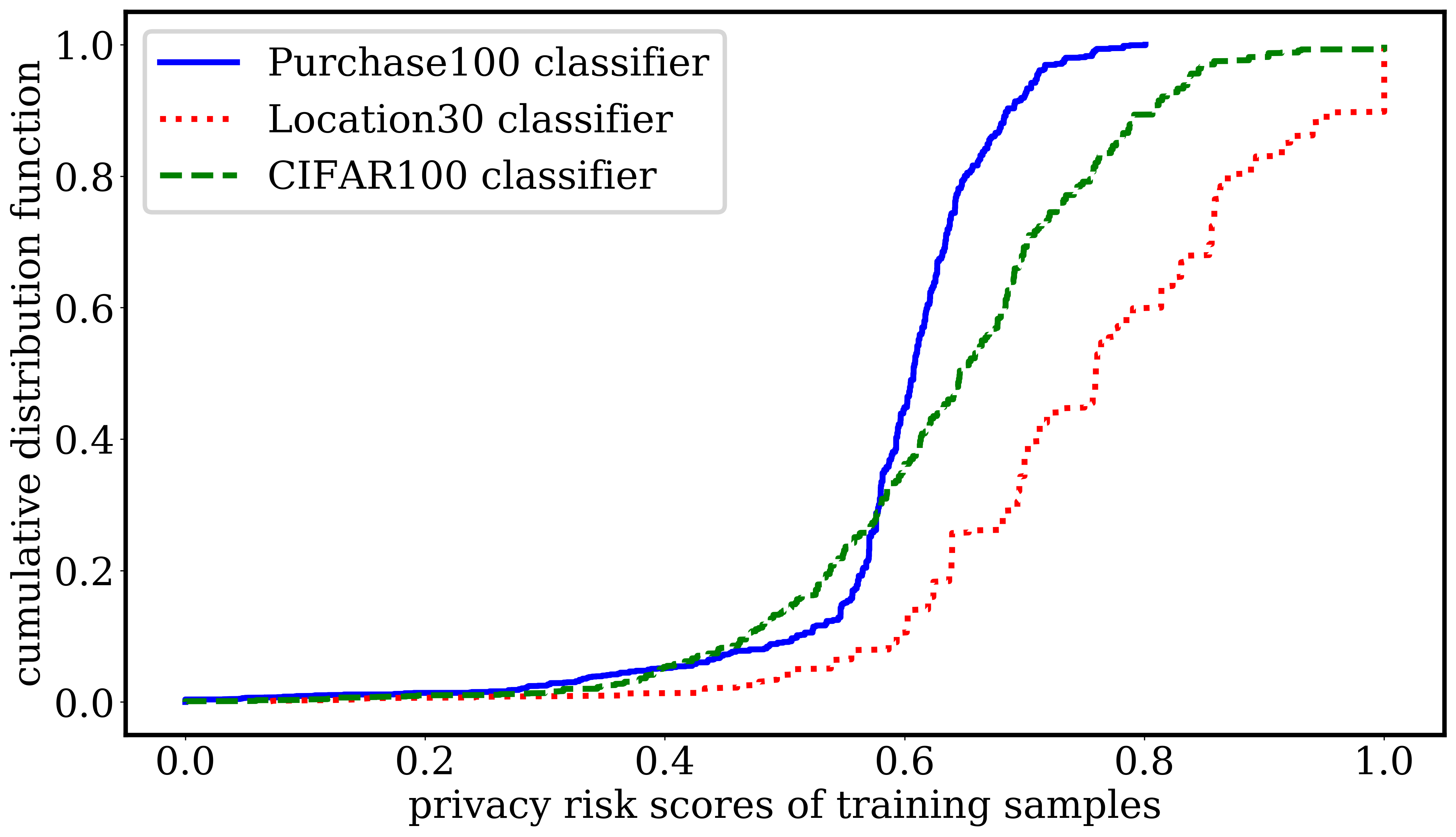}
	\caption{Cumulative distribution of privacy risk scores for undefended models trained on Purchase100, Location30, and CIFAR100 datasets.
	}\vspace{-3mm}
	\label{fig:risk_score_intro}
\end{figure}

To overcome the limitation of a lack of understanding of fine-grained privacy risks in existing works, we propose a new metric called the privacy risk score, that represents an \emph{individual} sample's \emph{probability} of being a member in the target model’s training set. 
Figure~\ref{fig:risk_score_intro} shows the cumulative distributions of privacy risk scores on target undefended models trained on Purchase100, Location30, and CIFAR100 datasets respectively.
We can see that the privacy risk faced by individual training samples is heterogeneous.
By utilizing the privacy risk score, an adversary can perform membership inference attacks with \emph{high confidence}: an input sample is inferred as a member if and only if its privacy risk score is higher than a certain threshold value.
Overall, we recommend that our per-sample privacy risk analysis should be used in conjunction with existing aggregate privacy analysis for an in-depth understanding of privacy risks of machine learning models.
Conventional aggregate analysis provides an average perspective of privacy risks incurred by all samples, while privacy risk score provides a perspective on privacy risk from the viewpoint of an individual sample. The former provides an aggregate estimation of privacy risks, while the latter allows us to understand the \emph{heterogeneous distribution} of privacy risks faced by individual samples and identify samples with high privacy risks. 
We summarize our contributions as follows:
\begin{enumerate}
    \item 
    We propose a suite of non-NN based attacks to benchmark target models' privacy risks by improving existing attacks with class-specific threshold settings and designing a new inference attack based on a modified prediction entropy estimation in a manner that incorporates the ground truth class label. 
    Furthermore, to rigorously evaluate the performance of membership inference defenses, we make recommendations for comparison with early stopping baseline and considering adaptive attackers with knowledge of defense mechanisms. 
    
    \item With our benchmark attacks, we find that two state-of-the-art defense approaches \cite{nasr_membership_defense_CCS18, jia2019memguard_ccs19} 
    are not as effective as previously reported. 
    Furthermore, we observe that the defense performance of adversarial regularization \cite{nasr_membership_defense_CCS18} is no better than early stopping, and the evaluation of MemGuard \cite{jia2019memguard_ccs19} lacks a consideration of adaptive adversaries.
    We also find that the existing white-box attacks \cite{nasr_whitebox_privacy_SP19} have limited advantages over our benchmark attacks, which only need black-box access to the target model.
    We also show that our attacks with class-specific threshold settings strictly outperform attacks with class-independent thresholds, and our new inference attack based on modified prediction entropy strictly outperforms conventional prediction entropy based attack.

    \item We propose to analyze privacy risks of machine learning models in a fine-grained manner by focusing on individual samples. We define a new metric called the privacy risk score, that estimates an \emph{individual} sample’s probability of being in the target model’s training set. 
    \item We experimentally demonstrate the effectiveness of our new metric in being able to capture the likelihood of an individual sample being a training member. We also show how an adversary can exploit our metric to launch membership inference attacks on individual samples with high confidence. 
    Finally we perform an in-depth investigation of our privacy risk score metric, and its correlations with model sensitivity, generalization error, and feature embeddings.
\end{enumerate}

Our code is publicly available at \url{https://github.com/inspire-group/membership-inference-evaluation} for the purpose of reproducible research. Furthermore, our evaluation mechanisms have also been integrated in Google's TensorFlow Privacy library. 

\section{Background and Related Work}
In this section, we first briefly introduce machine learning basics and notations.
Next, we present existing membership inference attacks, including black-box attacks and white-box attacks.
Finally, we discuss two state-of-the-art
defense methods: adversarial regularization \cite{nasr_membership_defense_CCS18} and MemGuard \cite{jia2019memguard_ccs19}.

\subsection{Machine learning basics and notations}
Let $F_{\theta}: \mathds{R}^{d} \rightarrow \mathds{R}^{k}$ be a machine learning model with $d$ input features and $k$ output classes, parameterized by weights $\theta$.
For an example $\bfz = (\bfx, y)$ with the input feature $\bfx$ and the ground truth label $y$, 
the model outputs a prediction vector over all class labels $F_{\theta}(\bfx)$ with $\sum_{i=0}^{k-1} F_{\theta}(\bfx)_{i} = 1$, 
and the final classification result will be the label with the largest prediction probability $\hat{y} = \argmax_{i} F_{\theta}(\bfx)_{i}$.

Given a training set $D_{\text{tr}}$, the model weights are optimized
by minimizing the prediction loss over all training examples.
\begin{equation}\label{eq:nat-train}
\min_{\theta} \, \frac{1}{|D_{\text{tr}}|}\sum_{\bfz \in D_{\text{tr}}} \mathcal{\ell}(F_{\theta},\bfz),
\end{equation}
where $|D_{\text{tr}}|$ denotes the size of training set, and $\ell$ computes the prediction loss, such as cross-entropy loss. 
In this paper, we skip the model parameter $\theta$ for simplicity and use $F$ to denote the machine learning model.

\subsection{Membership inference attacks}

For a target machine learning model, membership inference attacks aim to determine whether a given data point was used to train the model or not 
\cite{shokri_membership_SP17, yeom_privacy_CSF18, salem_membership_NDSS19, long_mem_arxiv18}. 
The attack poses a serious privacy risk to the individuals whose data is used for model training, for example in the setting of health analytics.

\subsubsection{Black-box membership inference attacks}
Shokri et al. \cite{shokri_membership_SP17} investigated the membership inference attacks against machine learning models in the black-box setting.
For an input sample $\bfz=(\bfx, y)$ to the target model $F$, the adversary only observes the prediction output $F(\bfx)$ and infers if $\bfz$ belongs to the model's training set $D_{\text{tr}}$.
To distinguish between target model's predictions on members and non-members, the adversary learns an attack model using the shadow training technique:
(1) the adversary first trains multiple shadow models to simulate the behavior of the target model;
(2) based on shadow models' outputs on their own training and test examples, the adversary obtains a labeled (member vs non-member) dataset, and 
(3) finally trains multiple neural network (NN) classifiers, one for each class label, to perform inference attacks against the target model.

Salem et al. \cite{salem_membership_NDSS19} show that even with only a single shadow model, membership inference attacks are still quite successful.
Furthermore, in the case where the adversary knows a subset of target model's training set and test set, the attack classifier can be directly trained with target model's predictions on those known samples, and then tested on unknown training and test sample \cite{nasr_membership_defense_CCS18, nasr_whitebox_privacy_SP19}.
Nasr et al. \cite{nasr_membership_defense_CCS18} redesign the attack by using one-hot encoded class labels as part of input features and training a single NN attack classifier for all class labels.

Besides membership inference attacks that rely on training NN classifiers, there are non-NN based attack methods that make inference decisions based on computing custom metrics on the predictions of the target model.
Leino et al. \cite{leino_mem_arxiv19} suggest using the metric of prediction correctness as a sign of being a member or not.
Yeom et al. \cite{yeom_privacy_CSF18} and Song et al. \cite{song_privacyriks_CCS19} find that the metric of prediction confidence of correct label $F(\bfx)_{y}$ can be compared with a certain threshold value to achieve similar attack performance as NN-based attacks.
Shokri et al. \cite{shokri_membership_SP17} show a large divergence between prediction entropy distributions over training data and test data, although this metric was not explicitly used for attacks.

Despite the existence of such non-NN based attacks, many research papers \cite{nasr_membership_defense_CCS18, nasr_whitebox_privacy_SP19, jia2019memguard_ccs19} still only train NN attack classifiers to evaluate target models' privacy risks.
We find that this can lead to severe underestimation of privacy risks by re-evaluating the same target models with non-NN based attacks. 
Furthermore, we improve existing non-NN based attacks by setting different threshold values for different class labels, building upon the motivation of separated attack classifiers for each class label by Shokri et al. \cite{shokri_membership_SP17}.
We also propose a new inference attack method by considering ground truth label when evaluating prediction uncertainty.

\subsubsection{White-box membership inference attacks}
Nasr et al. \cite{nasr_whitebox_privacy_SP19} analyze membership inference attacks in the white-box setting, where the adversary has the full access to the target machine learning model and knows the model architecture and model parameters.
They find that simply combining target model's final predictions and its intermediate computations to learn the attack classifier results in attack accuracy no better than that of the corresponding black-box attacks.
Instead, by using the gradient of prediction loss with regard to model parameters $\frac{\partial{\mathcal{\ell}(F_{\theta},\bfz)}}{\partial{\theta}}$ as additional features, the white-box membership inference attacks obtain higher attack accuracy than the black-box attacks.
We show that the gap between white-box attack accuracy and black-box attack accuracy is much smaller than previous estimates in this paper.

\subsection{Defenses against membership inference attacks}
To mitigate the risks of membership inference attacks, several defense ideas have been proposed.
$L_2$ norm regularization \cite{weight_decay_krogh_nips1992} and dropout \cite{srivastava_dropout_JMLR14} are standard techniques for reducing overfitting in machine learning.
They are also shown to decrease privacy risks to some degree \cite{shokri_membership_SP17, salem_membership_NDSS19}.
However, target models can still be quite vulnerable after applying these techniques.
Differential privacy \cite{Dwork_DP_first, DP_paper_wiki} can also be applied to ML models for provable risk mitigation \cite{abadi_DP_DL_CCS16, shokri2015privacy, pate_papernot_iclr17, mcmahan_DP_RNN_iclr18}, however, it induces significant accuracy drop for desired values of the privacy parameter 
\cite{jayaraman_evaluate_dp_usenix19}.
Two dedicated defenses, adversarial regularization \cite{nasr_whitebox_privacy_SP19} and MemGuard \cite{jia2019memguard_ccs19}, were recently proposed against membership inference attacks.
Both defenses are reported to have the ability of decreasing the attack accuracy to around $50\%$, which is the performance of random guessing.
We explain their details below.

\subsubsection{Adversarial regularization \cite{nasr_membership_defense_CCS18}}
Nasr et al. \cite{nasr_membership_defense_CCS18} propose to include the membership inference adversary with the NN-based attack into the training process itself to mitigate privacy risks.
At each training step, the attack classifier is first updated to distinguish between training data (members) and validation data (non-members), and then the target classifier is updated to simultaneously minimize the prediction loss and mislead the attack classifier.

More specifically, to train the classifier $F$ with parameters $\theta$ in a manner that is resilient against  membership inference attacks, Nasr et al. \cite{nasr_membership_defense_CCS18} use another classifier $I$ with parameters $\vartheta$ to perform membership inference attacks.
The attack classifier $I$ takes the target model's prediction $F(\bfx)$ and the input label $y$ as input features and generate one single output $I(F(\bfx), y)$, which is in the range [0, 1].
It infers the input sample as a member if the output is larger than 0.5, a non-member otherwise.
At each training step, they first update the attack classifier $I$ by maximizing the membership inference gain over the training set $D_{\text{tr}}$ and the validation set $D_{\text{val}}$.
\begin{equation}\label{eq:adv_regularize1}
\argmax_{\vartheta} \frac{\sum_{\bfz \in D_{\text{tr}}} \log(I(F(\bfx), y))}{2|D_{\text{tr}}|} 
+ \frac{\sum_{\bfz \in D_{\text{val}}} \log(1-I(F(\bfx), y))}{2|D_{\text{val}}|}  \\
\end{equation}
They further train the target classifier by minimizing both model prediction loss and membership inference gain over the training set $D_{\text{tr}}$.
\begin{equation}\label{eq:adv_regularize2}
\argmin_{\theta} \frac{1}{|D_{\text{tr}}|}\sum_{\bfz \in D_{\text{tr}}} \ell(F(\bfx),y) + \lambda \log(I(F(\bfx), y)),
\end{equation}
where $\lambda$ is a penalty parameter for the privacy risk.
In this way, the target model $F$ is trained with an additional regularization term to defend against membership inference attacks.

\subsubsection{MemGuard \cite{jia2019memguard_ccs19}}
Jia et al. \cite{jia2019memguard_ccs19} propose MemGuard as a defense method against membership inference attacks, which, different from Nasr et al. \cite{nasr_membership_defense_CCS18}, does not need to modify the training process.
Instead, given a pre-trained target model $F$, they obfuscate its predictions with well-designed noises to confuse the membership inference classifier $I$, without changing classification results.

The attack classifier $I$ is trained following the shadow-training technique \cite{shokri_membership_SP17}, which takes the model prediction $F(\bfx)$ with the sample label $y$, and outputs a score $I(F(\bfx), y)$ in the range [0 ,1] for membership inference: if the output is larger than 0.5, the data sample is inferred as a member, and vice versa.
The key question of how to add noise $\bfn$ to $F(\bfx)$ can be formulated as the following optimization problem:
\begin{equation}\label{eq:mem_guard}
\begin{aligned}
& \min_{\bfn}  d(F(\bfx)+\bfn, F(\bfx)),\\
\text{subject to:} & \argmax_{i} (F(\bfx)_{i}+\bfn_{i}) =  \argmax_{i} F(\bfx)_{i}, \\
& I(F(\bfx)+\bfn) = 0.5, \\
& F(\bfx)_{i}+\bfn_{i} \geq 0 , \forall i\\
& \sum_{i}\bfn_{i} = 0,
\end{aligned}
\end{equation}
where the objective is to minimize the distance $d$ between original predictions and noisy predictions. 
The first constraint ensures the classification result does not change after adding noise, the second constraint ensures the attack classifier cannot determine whether the sample is a member or a non-member with the noisy predictions, and last two constraints ensure the noisy predictions are valid.

When evaluating the defense performance, both Nasr et al. \cite{nasr_membership_defense_CCS18} and Jia et al. \cite{jia2019memguard_ccs19} train NN classifiers for inference attacks.
As shown in the following section, we find that their evaluations underestimate privacy risks.
With our benchmark attacks, the adversary achieves significantly higher attack accuracy on defended models than previous estimates. 
We further find that the performance of adversarial regularization \cite{nasr_membership_defense_CCS18} is no better than  early stopping, and the evaluation of MemGuard \cite{jia2019memguard_ccs19} lacks consideration of strategic adversaries.

\section{Systematically Evaluating Membership Inference Privacy Risks}\label{sec:benchmarks}

In this section, we first present a suite of non-NN based attacks to benchmark privacy risks, which only need to observe target model's output predictions (i.e., black-box setting).
Next, we provide two recommendations, comparison with early stopping and considering adaptive attacks, to rigorously measure the effectiveness of defense approaches.
Finally, we present experiment results by re-evaluating target models in prior work \cite{nasr_membership_defense_CCS18, jia2019memguard_ccs19, nasr_whitebox_privacy_SP19} with our proposed benchmark attacks.

\subsection{Benchmarks of membership inference attacks}\label{subsec:benchmark_attacks}

We propose to use a suite of non-NN based attack methods to benchmark membership inference privacy risks of machine learning models.
We call these attack methods \emph{``metric-based attacks''} as they first measure the performance metrics of target model's predictions, such as correctness, confidence, and entropy, and then compare those metrics with certain threshold values to infer whether the input sample is a member or a non-member \cite{song_privacyriks_CCS19, leino_mem_arxiv19}.
We improve existing metric-based attacks by setting different threshold values for different class labels of target models.
Then we propose another new metric-based attack by considering ground truth label when evaluating prediction uncertainty.
We denote the inference strategy as $\mathcal{I}$, which codes members as 1, and non-members as 0.
Overall, we propose that existing NN based attacks should be supplemented with our metric-based attacks for systematically and rigorously evaluating privacy risks of ML models.

\subsubsection{Existing attacks}

\noindent \textbf{Inference attack based on prediction correctness}
Leino et al. \cite{leino_mem_arxiv19} observe that the membership inference attacks based on whether the input is classified correctly or not achieve comparable success as NN-based attack on target models with large generalization errors.
The intuition is that the target model is trained to predict correctly on training data (members), which may not generalize well on test data (non-members).
Thus, we can rely on the prediction correctness metric for membership inference.
The adversary infers an input sample as a member if it is correctly predicted, a non-member otherwise.
\begin{equation}\label{eq:corr-based-infer}
\mathcal{I}_{\text{corr}}(F, (\bfx, y)) = \mathds{1}\{ \argmax_{i}F(\bfx)_{i} = y \},
\end{equation}
where $\mathds{1}\{\cdot\}$ is the indicator function.

\subsubsection{Improving existing attacks with class-dependent thresholds}\label{subsubsec:improved_attacks}

\noindent \textbf{Inference attack based on prediction confidence}
Yeom et al. \cite{yeom_privacy_CSF18} and Song et al. \cite{song_privacyriks_CCS19} show that 
the attack strategy of using a threshold on the prediction confidence results in similar attack accuracy as NN-based attacks.
The intuition is that the target model is trained by minimizing prediction loss over training data, which means the prediction confidence of a training sample $F(\bfx)_{y}$ should be close to 1.
On the other hand, the model is usually less confident in predictions on a test sample.
Thus, we can rely on the metric of prediction confidence for membership inference.
The adversary infers an input example as a member if its prediction confidence is larger than a preset threshold, a non-member otherwise.
\begin{equation}\label{eq:conf-based-infer}
\mathcal{I}_{\text{conf}}(F, (\bfx, y)) = \mathds{1}\{ F(\bfx)_{y} \geq \tau_{y} \}.
\end{equation}
Yeom et al. \cite{yeom_privacy_CSF18} and Song et al. \cite{song_privacyriks_CCS19} choose to use a single threshold for all class labels.
We improve this method by setting different threshold values $\tau_{y}$ for different class labels $y$. 
The reason is that the dataset may be unbalanced so that the target model indeed has different confidence levels for different class labels.
Our experiments show that this class-dependent thresholding technique leads to better attack performance.
The class-dependent threshold values $\tau_{y}$ are learned with the shadow-training technique \cite{shokri_membership_SP17}: the adversary (1) first trains a shadow model to simulate the behavior of the target model; (2) then obtains the shadow model's prediction confidence values on both shadow training and shadow test data; (3) finally 
leverages knowledge of membership labels (member vs non-member) of the shadow data to 
select the threshold value $\tau_{y}$ which achieves the highest accuracy in distinguishing between shadow training data and shadow test data with the class label $y$ based on Equation \eqref{eq:conf-based-infer}.

\noindent \textbf{Inference attack based on prediction entropy}
Although there is no prior work using prediction entropy for membership inference attacks, 
Shokri et al. \cite{shokri_membership_SP17} indeed present the difference of prediction entropy distributions between training and test data to explain why privacy risks exist.
Salem et al. \cite{salem_membership_NDSS19} also mention the possibility of using prediction entropy for attacks.
The intuition is that the target model is trained by minimizing the prediction loss over training data, which means the prediction output of a training sample should be close to a one-hot encoded vector and its prediction entropy should be close to 0.
On the other hand, the target model usually has a larger prediction entropy on an unseen test sample.
Therefore, we can rely on the metric of prediction entropy for membership inference.
The adversary classifies an input example as a member if its prediction entropy is smaller than a preset threshold, a non-member otherwise.
\begin{equation}\label{eq:entr-based-infer}
\mathcal{I}_{\text{entr}}(F, (\bfx, y)) = \mathds{1}\{ -\sum_{i}F(\bfx)_{i} \log(F(\bfx)_{i}) \leq \hat{\tau}_{y} \}.
\end{equation}
Similar to the confidence-based attack, we propose to use the threshold values $\hat{\tau}_{y}$ that are dependent on the class labels and are set with the shadow-training technique \cite{shokri_membership_SP17}.

\subsubsection{Our new inference attack based on modified prediction entropy}

The attack based on prediction entropy has one serious issue: it does not contain any information about the ground truth label. 
In fact, both a correct classification with probability of 1 and a totally wrong classification with probability of 1 lead to zero prediction entropy values.

To resolve this issue, we design a new metric with following two properties to measure the model prediction uncertainty given the ground truth label: it should be (1) monotonically decreasing with the prediction probability of the correct label $F(\bfx)_y$, and (2) monotonically increasing with the prediction probability of any incorrect label $F(\bfx)_{i}, \forall i \neq y$.
Let $x \in [0,1]$ denote the prediction probability for a certain label, the function used in conventional entropy computations $-x \log x$ is not a monotonic function. As a contrast, $-(1-x)\log x$ is a monotonically decreasing function, and $-x\log(1-x)$ is a monotonically increasing function.
Therefore, we propose the modified prediction entropy metric, computed as follows.
\begin{equation}\label{eq:m-entr-def}
\begin{aligned}
\mentropy(F(\bfx), y) = & -(1-F(\bfx)_{y}) \log(F(\bfx)_{y}) \\
 & - \sum_{i \neq y} F(\bfx)_{i} \log(1-F(\bfx)_{i}).
\end{aligned}
\end{equation}
In this way, a correct classification with probability of 1 leads to modified entropy of 0, while a wrong classification with probability of 1 leads to modified entropy of infinity.

Now, with the new metric of modified prediction entropy, the adversary classifies an input example as a member if its modified prediction entropy is smaller than a preset threshold, a non-member otherwise.
\begin{equation}\label{eq:m-entr-based-infer}
\mathcal{I}_{\mentropy}(F, (\bfx, y)) = \mathds{1}\{ \mentropy(F(\bfx), y)  \leq \check{\tau}_{y} \}.
\end{equation}
Similar to previous scenarios, we set different threshold values $\check{\tau}_{y}$ for different class labels, which are learned with the shadow training technique \cite{shokri_membership_SP17}.
Experiments show that the inference attack based on our modified prediction entropy is strictly superior to the inference attack based on prediction entropy.

\subsection{Rigorously evaluating membership inference defenses}
To evaluate the effectiveness of defenses against membership inference attacks, we make the following two recommendations, besides using our metric-based benchmark attacks.

\subsubsection{Comparison with early stopping}
\label{subsub:early_stopping}
During the training process, the target model's parameters are updated following gradient descent methods, so the training error and test error usually get reduced gradually with an increasing number of training epochs.
However, as the number of training epochs increases, the target model also becomes more vulnerable to membership inference attacks, due to increased memorization. 
We thus propose \emph{early stopping}~\cite{early_stopping_prechelt1998, caruana2001overfitting, early_stopping_yao2007} as a benchmark defense method, in which fewer training epochs are used in order to tradeoff a slight reduction in model accuracy with lower privacy risk.  

\begin{figure}[!ht]
	\centering
	\includegraphics[width=\linewidth]{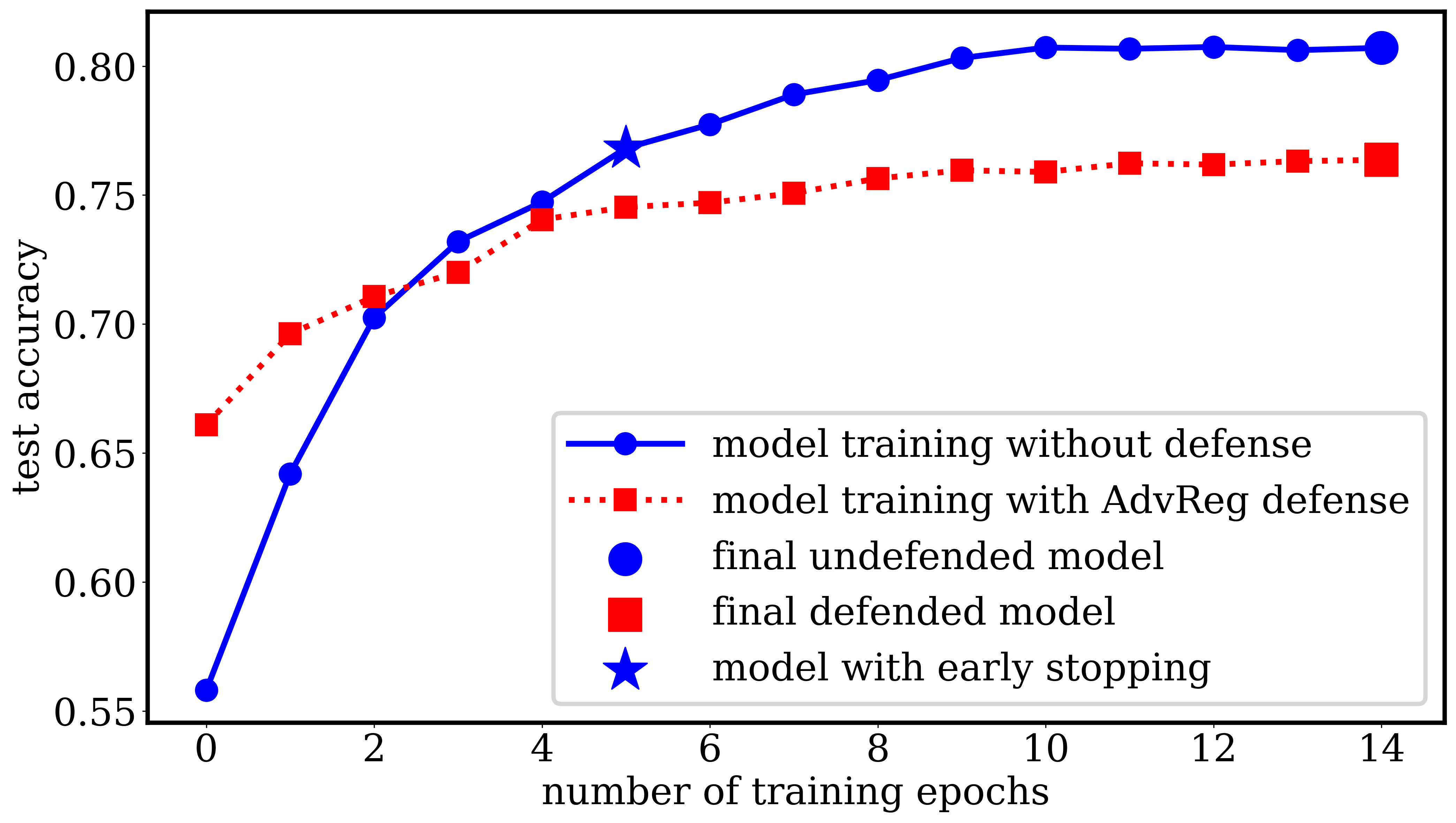}
	\caption{
	Test accuracy at different training epochs for Purchase100 classifiers without defense and with adversarial regularization defense \cite{nasr_membership_defense_CCS18}.
	We should compare the final defended model to the model with early stopping.
	}
	\label{fig:acc_plot}
\end{figure}

We recommend that whenever a defense method is proposed in the literature that reduces the threat of membership inference attacks at the cost of degradation in model accuracy, the performance of the defense method should be benchmarked against our early stopping approach. This is indeed the case for the defense method of adversarial regularization (AdvReg) \cite{nasr_membership_defense_CCS18}. 
As shown in Figure~\ref{fig:acc_plot}, the defended Purchase100 classifier should be compared to the undefended model with fewer training steps and similar accuracy.

\subsubsection{Adaptive attacks}\label{subsub:adaptive_attacks}

There always exists an arms race between privacy attacks and defenses for machine learning models.
When evaluating the defense performance, it is critical to put the adversary into the last step, i.e., the adversary knows the defense mechanism and performs adaptive attacks against the defended models.
A perfect defense performance with non-adaptive attacks does not mean that the defense approach is effective
\cite{carlini_bypass_defense_AIsec17, he2017adversarial, athalye_adv_ICML18}.

Specifically for defenses proposed against membership inference attacks, we should consider that the adversary knows the defense mechanism such that he or she can train shadow models following the defense method.
From these defended shadow models, the adversary then learns an attack classifier or sets threshold values for metric-based attacks, and finally performs attacks on the defended target model.

\subsection{Experiment results} \label{subsec:MIA_results}

We first re-evaluate the effectiveness of two membership inference defenses \cite{nasr_membership_defense_CCS18, jia2019memguard_ccs19}, and then re-evaluate the white-box membership inference attacks proposed by Nasr et al. \cite{nasr_whitebox_privacy_SP19}.
Following prior work \cite{shokri_membership_SP17, yeom_privacy_CSF18, song_privacyriks_CCS19}, we sample the input $(\bfx,y)$ from either the target model's training set or test set with an equal $0.5$ probability to maximize the uncertainty of membership inference attacks. 
Thus, the random guessing strategy results in a $50\%$ membership inference attack accuracy.

\subsubsection{Datasets}
\noindent \textbf{Purchase100} This dataset is based on Kaggle's 
Acquire Valued Shoppers Challenge,\footnote{\url{https://www.kaggle.com/c/acquire-valued-shoppers-challenge}} which contains shopping records of several thousand individuals.
We obtain a simplified and preprocessed purchase dataset provided by Shokri et al. \cite{shokri_membership_SP17}.
The dataset has 197,324 data samples with 600 binary features. Each feature corresponds to a product and represents whether the individual has purchased it or not.
All data samples are clustered into 100 classes representing different purchase styles.
The classification task is to predict the purchase style based on the 600 binary features.
We follow Nasr et al. \cite{nasr_membership_defense_CCS18, nasr_whitebox_privacy_SP19} to use 10\% data samples (19,732) to train a model.

\noindent \textbf{Texas100}
This dataset is based on the Hospital Discharge Data public use files with patients' information released by the Texas Department of State Health Services.\footnote{\url{https://www.dshs.texas.gov/THCIC/Hospitals/Download.shtm}} 
Each data record contains the external causes of injury (e.g., suicide, drug misuse), the diagnosis (e.g., schizophrenia), the procedures the patient underwent (e.g., surgery) and some generic information (e.g., gender, age, race).
We obtain a simplified and preprocessed Texas dataset provided by Shokri et al. \cite{shokri_membership_SP17}.
The classification task is to predict the patient's main procedure based on the patient's information.
The dataset focuses on 100 most frequent procedures, resulting in 67,330 data samples with 6,170 binary features.
Following previous papers \cite{nasr_membership_defense_CCS18, nasr_whitebox_privacy_SP19, jia2019memguard_ccs19}, we use 10,000 data samples to train a model.

\noindent \textbf{Location30}
This dataset is based on Foursquare dataset,\footnote{\url{https://sites.google.com/site/yangdingqi/home/foursquare-dataset}} which contains location ``check-in'' records of several thousand individuals.
We obtain a simplified and preprocessed Location dataset provided by Shokri et al. \cite{shokri_membership_SP17}.
The dataset contains 5,010 data samples with with 446 binary features.
Each feature corresponds to a certain region or location type and represents whether the individual has visited the region/location or not.
All data samples are clustered into 30 classes representing different geosocial types.
The classification task is to predict the geosocial type based on the 466 binary features.
Following Jia et al. \cite{jia2019memguard_ccs19}, we use 1,000 data samples to train a model.

\noindent \textbf{CIFAR100}
This is a major benchmark dataset for image classification \cite{cifar_krizhevsky2009learning}.
It is composed of 32$\times$32 color images in 100 classes, with 600 images per class.
For each class label, 500 images are used as training samples, and the remaining 100 images are used as test samples.

We choose these datasets 
for fair comparison with prior work \cite{nasr_membership_defense_CCS18, nasr_whitebox_privacy_SP19, jia2019memguard_ccs19}. Since all datasets except CIFAR100 are binary datasets, we also provide attack results with more complex datasets in Appendix \ref{appendix:attacks_other_data}, where our benchmark attacks achieve higher attack success than NN-based attacks.

\begin{table*}[!ht]
\caption{Benchmarking the effectiveness of using adversary regularization \cite{nasr_membership_defense_CCS18} as defense against membership inference attacks.
We can see that the defended models are still vulnerable to membership inference attacks.
}\vspace{-0.5em}
\centering
\renewcommand\arraystretch{1.0}
\fontsize{9pt}{9pt}\selectfont
\begin{tabular}{cccc|ccccc}
\toprule[1.5pt]
\multicolumn{4}{c|}{\textbf{\normalsize{Model Performance}}} & \multicolumn{5}{c}{\textbf{\normalsize{Membership Inference Attacks}}} \\
\multirow{2}{*}{{dataset}} & {using} & {training} & {test} & {attack acc}  & {attack acc}  & {attack acc}  & {attack acc} & {attack acc}   \\
& {defense \cite{nasr_membership_defense_CCS18}?} & {acc} &  {acc} & { by \cite{nasr_membership_defense_CCS18}} & { ($\mathcal{I}_{\text{corr}}$)} & { ($\mathcal{I}_{\text{conf}}$)} & { ($\mathcal{I}_{\text{entr}}$)} & { ($\mathcal{I}_{\mentropy}$)} \\
\midrule[0.75pt]
\multirow{2}{*}{Purchase100} & \multirow{2}{*}{no} & \multirow{2}{*}{99.8\%} & \multirow{2}{*}{80.9\%} & \multirow{2}{*}{\textbf{67.6\%}} & \multirow{2}{*}{59.5\%} & \multirow{2}{*}{{67.1\%}} & \multirow{2}{*}{65.7\%} & \multirow{2}{*}{67.1\%}\\
 &  & & & & & & & \\
\multirow{2}{*}{Purchase100} & \multirow{2}{*}{yes} & \multirow{2}{*}{92.7\%} & \multirow{2}{*}{76.6\%} & \multirow{2}{*}{51.6\%} & \multirow{2}{*}{58.1\%} & \multirow{2}{*}{{59.4\%}} & \multirow{2}{*}{55.8\%} & \multirow{2}{*}{\textbf{59.5\%}}\\
& & & & & & & & \\
\multirow{2}{*}{Purchase100} & \multirow{2}{*}{early stopping} & \multirow{2}{*}{92.9\%} & \multirow{2}{*}{76.4\%} & \multirow{2}{*}{{N.A.}} & \multirow{2}{*}{58.2\%} & \multirow{2}{*}{\textbf{59.2\%}} & \multirow{2}{*}{55.9\%} & \multirow{2}{*}{59.1\%}\\
 &  & & & & & & & \\
\midrule[0.75pt]
\multirow{2}{*}{Texas100} & \multirow{2}{*}{no} & \multirow{2}{*}{81.0\%} & \multirow{2}{*}{52.3\%} & \multirow{2}{*}{{63.0\%}} & \multirow{2}{*}{64.4\%} & \multirow{2}{*}{\textbf{67.8\%}} & \multirow{2}{*}{60.2\%} & \multirow{2}{*}{67.7\%}\\
 &  & & & & & & & \\
\multirow{2}{*}{Texas100} & \multirow{2}{*}{yes} & \multirow{2}{*}{56.6\%} & \multirow{2}{*}{46.4\%} & \multirow{2}{*}{51.0\%} & \multirow{2}{*}{55.1\%} & \multirow{2}{*}{\textbf{58.6\%}} & \multirow{2}{*}{53.5\%} & \multirow{2}{*}{\textbf{58.6\%}}\\
& & & & & & & & \\
\multirow{2}{*}{Texas100} & \multirow{2}{*}{early stopping} & \multirow{2}{*}{59.3\%} & \multirow{2}{*}{47.9\%} & \multirow{2}{*}{{N.A.}} & \multirow{2}{*}{55.7\%} & \multirow{2}{*}{{59.4\%}} & \multirow{2}{*}{54.0\%} & \multirow{2}{*}{\textbf{59.5\%}}\\
 &  & & & & & & & \\
\bottomrule[1.5pt]
\end{tabular}
\label{tab:ccs18_defense}
\end{table*}

\subsubsection{Re-evaluating adversarial regularization \cite{nasr_membership_defense_CCS18}}
\label{subsubsec:ccs18}

We follow Nasr et al. \cite{nasr_membership_defense_CCS18} to train both defended and undefended classifiers on Purchase100 and Texas100 datasets.
For both datasets, the model architecture is a fully connected neural network with 4 hidden layers. The numbers of neurons for hidden layers are 1024, 512, 256, and 128, respectively. All hidden layers use hyperbolic tangent (Tanh) as the activation function.
We note that the defense method of adversarial regularization \cite{nasr_membership_defense_CCS18} incurs accuracy drop.
After applying the defense, the test accuracy drops from $80.9\%$ to $76.6\%$ on the Purchase100 dataset, and from $52.3\%$ to $46.4\%$ on the Texas100 dataset.
As we discuss in Section~\ref{subsub:early_stopping}, to further evaluate the effectiveness of adversarial regularization \cite{nasr_membership_defense_CCS18}, we also obtain models with early stopping by saving the \emph{undefended} models in every training epoch and picking the saved epochs with similar accuracy performance as defended models.
Table~\ref{tab:ccs18_defense} presents the membership inference attack results.

From Table~\ref{tab:ccs18_defense}, we can see that \textbf{the defended models are still vulnerable to membership inference attacks}, indicating the necessity of our metric-based benchmark attacks. We achieve $59.5\%$ and $58.6\%$ attack accuracy on the defended Purchase100 classifier and the defended Texas100 classifier with our benchmark attacks, significantly larger than $51.6\%$ and $51.0\%$ as reported by Nasr et al. \cite{nasr_membership_defense_CCS18}.
Furthermore, on all models except the undefended Purchase100 classifier, \textbf{the largest attack accuracy achieved by benchmark attacks is larger than that of NN based attacks} used in Nasr et al. \cite{nasr_membership_defense_CCS18}.
Note that the defense method provides limited mitigation of privacy risks: it reduces attack accuracy from around $67\%$ to around $59\%$ on tested models.
We also find that \textbf{our new attack based on the modified entropy ($\mathcal{I}_{\mentropy}$) always outperforms the conventional entropy based attack ($\mathcal{I}_{\text{entr}}$)}.
It is also very competitive among all benchmark attacks.

From Table~\ref{tab:ccs18_defense}, we also surprisingly find that \textbf{adversarial regularization \cite{nasr_membership_defense_CCS18} is no better than our early stopping benchmark method}: with early stopping, the undefended Purchase100 classifier and the undefended Texas100 classifier have the attack accuracy of $59.2\%$ and $59.5\%$, which are quite close to those of defended models.
Therefore, when evaluating the effectiveness of a future defense mechanism that trades lower model accuracy for lower membership inference risk, we argue to compare the defended model to the naturally trained model with early stopping for a fair comparison.
We emphasize that our early stopping baseline can be calibrated to achieve similar model accuracy as the defended model. In contrast, the adversarial regularization approach may have a model accuracy which is different from the defended model under consideration, and will thus not represent a fair comparison.

\begin{table}[!ht]
\caption{Comparing attack performance between conventional class-independent thresholding attacks and our class-dependent thresholding attacks.
}\vspace{-0.5em}
\centering
\renewcommand\arraystretch{1.1}
\fontsize{9pt}{9pt}\selectfont
\begin{tabular}{cccc}
\toprule[1.5pt]
\multirow{2}{*}{attack methods} & \multicolumn{3}{c}{defense methods for Texas100 classifier}\\
& {no defense} & {AdvReg \cite{nasr_membership_defense_CCS18}} & {early stopping} \\
 \midrule[0.75pt]
 $\mathcal{I}_{\text{conf}}$ & \multirow{2}{*}{64.7\%} & \multirow{2}{*}{55.5\%} & \multirow{2}{*}{55.8\%} \\
 (class-independent) & & & \\
 $\mathcal{I}_{\text{conf}}$ & \multirow{2}{*}{67.8\%} &  \multirow{2}{*}{58.6\%} & \multirow{2}{*}{59.4\%} \\
 (class-dependent) & & & \\
 \midrule[0.75pt]
  $\mathcal{I}_{\text{entr}}$ & \multirow{2}{*}{58.3\%} & \multirow{2}{*}{52.9\%} &  \multirow{2}{*}{53.2\%} \\
 (class-independent) & & & \\
 $\mathcal{I}_{\text{entr}}$ & \multirow{2}{*}{60.2\%} & \multirow{2}{*}{53.5\%} &  \multirow{2}{*}{54.0\%}\\
 (class-dependent) & & & \\
\midrule[0.75pt]
  $\mathcal{I}_{\mentropy}$ & \multirow{2}{*}{64.8\%} & \multirow{2}{*}{55.4\%} & \multirow{2}{*}{55.9\%} \\
 (class-independent) & & & \\
 $\mathcal{I}_{\mentropy}$ & \multirow{2}{*}{67.7\%} &  \multirow{2}{*}{58.6\%} &  \multirow{2}{*}{59.5\%} \\
 (class-dependent) & & & \\
\bottomrule[1.5pt]
\end{tabular}
\label{tab:ccs18_threhsold_setting}
\end{table}

To show the attack improvement yielded by our class-dependent thresholding technique, we compare with metric-based attacks when the same threshold is applied to all class labels.
Table~\ref{tab:ccs18_threhsold_setting} shows the results on Texas100 classifiers without defense, with AdvReg \cite{nasr_membership_defense_CCS18}, and with early stopping.
We can see that \textbf{with the class-dependent thresholding technique, we increase the attack accuracy by 1\% -- 4\%.}

\begin{table*}[!ht]
\caption{Benchmarking the effectiveness of using MemGuard \cite{jia2019memguard_ccs19} as defense against membership inference attacks.
We can see that the defended models are still vulnerable to membership inference attacks.
}\vspace{-0.5em}
\centering
\renewcommand\arraystretch{1.0}
\fontsize{9pt}{9pt}\selectfont
\begin{tabular}{cccc|ccccc}
\toprule[1.5pt]
\multicolumn{4}{c|}{\textbf{\normalsize{Model Performance}}} & \multicolumn{5}{c}{\textbf{\normalsize{Membership Inference Attacks}}} \\
\multirow{2}{*}{{dataset}} & {using} & {training} & {test} & {attack acc}  & {attack acc}  & {attack acc}  & {attack acc}  & {attack acc}    \\
& {defense \cite{jia2019memguard_ccs19}?} & {acc} &  {acc} &{ by \cite{jia2019memguard_ccs19}} & { ($\mathcal{I}_{\text{corr}}$)} & { ($\mathcal{I}_{\text{conf}}$)} & { ($\mathcal{I}_{\text{entr}}$)}  & { ($\mathcal{I}_{\mentropy}$)} \\
\midrule[0.75pt]
\multirow{2}{*}{Location30} & \multirow{2}{*}{no} & \multirow{2}{*}{100\%} & \multirow{2}{*}{60.7\%} & \multirow{2}{*}{\textbf{81.1\%}} & \multirow{2}{*}{68.7\%} & \multirow{2}{*}{76.3\%} & \multirow{2}{*}{61.6\%} & \multirow{2}{*}{78.1\%} \\
 &  & & & & & & & \\
\multirow{2}{*}{Location30} & \multirow{2}{*}{yes} & \multirow{2}{*}{100\%} & \multirow{2}{*}{60.7\%} & \multirow{2}{*}{50.1\%} & \multirow{2}{*}{68.7\%} & \multirow{2}{*}{\textbf{69.1\%}} & \multirow{2}{*}{52.1\%} & \multirow{2}{*}{68.8\%}\\
& & & & & & & & \\
\midrule[0.75pt]
\multirow{2}{*}{Texas100} & \multirow{2}{*}{no} & \multirow{2}{*}{99.95\%} & \multirow{2}{*}{51.77\%} & \multirow{2}{*}{{74.0\%}} & \multirow{2}{*}{74.2\%} & \multirow{2}{*}{{79.0\%}} & \multirow{2}{*}{66.6\%} & \multirow{2}{*}{\textbf{79.4\%}} \\
 &  & & & & & & & \\
\multirow{2}{*}{Texas100} & \multirow{2}{*}{yes} & \multirow{2}{*}{99.95\%} & \multirow{2}{*}{51.77\%} & \multirow{2}{*}{50.3\%} & \multirow{2}{*}{\textbf{74.2\%}} & \multirow{2}{*}{{74.1\%}} & \multirow{2}{*}{54.6\%} & \multirow{2}{*}{74.0\%}\\
& & & & & & & & \\
\bottomrule[1.5pt]
\end{tabular}
\label{tab:ccs19_defense}
\end{table*}

\begin{table*}[!ht]
\caption{Benchmarking the effectiveness of white-box membership inference attacks proposed by Nasr et al. \cite{nasr_whitebox_privacy_SP19}.
We can see that compared with our black-box benchmark attacks, the advantage of white-box attacks is limited.
}\vspace{-0.5em}
\centering
\renewcommand\arraystretch{1.0}
\fontsize{9pt}{9pt}\selectfont
\begin{tabular}{ccc|c|ccccc}
\toprule[1.5pt]
\multicolumn{3}{c|}{\textbf{\normalsize{Model Performance}}} & \multicolumn{6}{c}{\textbf{\normalsize{Membership Inference Attacks}}} \\
\multirow{2}{*}{{dataset}} & {training} & {test} & {attack acc}  & {attack acc}  & {attack acc}  & {attack acc} & {attack acc} & {attack acc} \\
&   {acc} &  {acc} &{by \cite{nasr_whitebox_privacy_SP19} (white-box)} &{by \cite{nasr_whitebox_privacy_SP19} (black-box)} & { ($\mathcal{I}_{\text{corr}}$)} & { ($\mathcal{I}_{\text{conf}}$)} & { ($\mathcal{I}_{\text{entr}}$)} & { ($\mathcal{I}_{\mentropy}$)}\\
\midrule[0.75pt]
\multirow{2}{*}{Purchase100}  & \multirow{2}{*}{99.8\%} & \multirow{2}{*}{80.9\%} & \multirow{2}{*}{\textbf{73.4\%}} & \multirow{2}{*}{\textbf{67.6\%}} & \multirow{2}{*}{59.5\%} & \multirow{2}{*}{{67.1\%}} & \multirow{2}{*}{65.7\%}& \multirow{2}{*}{67.1\%}\\
 &  & & & & & & & \\
\midrule[0.75pt]
\multirow{2}{*}{Texas100}  & \multirow{2}{*}{81.0\%} & \multirow{2}{*}{52.3\%} & \multirow{2}{*}{\textbf{68.3\%}} & \multirow{2}{*}{{63.0\%}} & \multirow{2}{*}{64.4\%} & \multirow{2}{*}{\textbf{67.8\%}} & \multirow{2}{*}{60.2\%}& \multirow{2}{*}{67.7\%}\\
 &  & & & & & & & \\
\midrule[0.75pt]
\multirow{2}{*}{CIFAR100}  & \multirow{2}{*}{100\%} & \multirow{2}{*}{83.00\%} & \multirow{2}{*}{\textbf{74.3\%}} & \multirow{2}{*}{{67.7\%}} & \multirow{2}{*}{58.5\%} & \multirow{2}{*}{\textbf{73.7\%}} & \multirow{2}{*}{73.3\%} & \multirow{2}{*}{73.6\%}\\
 &  & & & & & & & \\
\bottomrule[1.5pt]
\end{tabular}
\label{tab:sp19_whitebox}
\end{table*}

\subsubsection{Re-evaluating MemGuard \cite{jia2019memguard_ccs19}}
\label{subsubsec:ccs19}
We follow Jia et al. \cite{jia2019memguard_ccs19} to train classifiers on Location30 and Texas100 datasets.
For both datasets, the model architecture is a fully connected neural network with 4 hidden layers. The numbers of neurons for hidden layers are 1024, 512, 256, and 128, respectively. All hidden layers use rectified linear unit (ReLU) as the activation function.
MemGuard \cite{jia2019memguard_ccs19} does not change the accuracy performance, so the comparison with early stopping benchmark is not applicable.
Table~\ref{tab:ccs19_defense} lists the attack accuracy on both undefended and defended models, 
with attack methods in Jia et al. \cite{jia2019memguard_ccs19} and our metric-based benchmark attack methods.
In fact, Jia et al. \cite{jia2019memguard_ccs19} use 6 different NN attack classifiers to measure the privacy risks, and we pick the highest attack accuracy among them.

From Table~\ref{tab:ccs19_defense}, we again emphasize the necessity of our benchmark attacks by showing that \textbf{the defended models still have high membership inference accuracy}: $69.1\%$ on the defended Location30 classifier and $74.2\%$ on the defended Texas100 classifier, much larger than $50.1\%$ and $50.3\%$ reported by Jia et al. \cite{jia2019memguard_ccs19}.
We even achieve \textbf{higher membership inference accuracy than attacks in Jia et al. \cite{jia2019memguard_ccs19}} on all models, except the undefended Location30 classifier.
Note that the defense still works but to a limited degree: it reduces the attack accuracy by $12\%$ on the Location30 classifier and by $5\%$ on the Texas100 classifier.
Similar to Section~\ref{subsubsec:ccs18}, \textbf{our proposed modified-entropy based attack always achieves higher attack accuracy than the entropy based attack, and is very competitive among all benchmark attacks.}

Next, we discuss why Jia et al. \cite{jia2019memguard_ccs19} fail to achieve high membership inference accuracy for their defended models. 
We find that most of their attacks (4 out of 6) are non-adaptive attacks, where the adversary has no idea of the implemented defense, and thus the membership inference attacks are not successful.
For the two adaptive attacks, Jia et al. \cite{jia2019memguard_ccs19} \textbf{do not put the adversary in the last step of the arms race between attacks and defenses}.
In their attacks, the adversary is aware that the model predictions will be perturbed with noises but does not know the exact algorithm of noise generation implemented by the defender.
In their first adaptive attack, Jia et al. \cite{jia2019memguard_ccs19} round the model predictions to be one decimal during the attack classifier's inference to mitigate the effect of the perturbation. However, the attack performance is greatly degraded when the applied perturbation is large.
In the second adaptive attack, Jia et al. \cite{jia2019memguard_ccs19} train the attack classifier using the state-of-the-art robust training algorithm by Madry et al. \cite{madry_robust_ICLR18}, with the hope that noisy perturbation will not change the classification.
However, the robust training algorithm \cite{madry_robust_ICLR18} has a very poor generalization property: the predictions on test points are still likely to be wrong after adding well-designed noises.
For a thorough evaluation of the defense, we should consider that the attacker has the full knowledge of the defense mechanism, and he or she learns the attack model based on the defended shadow models.

\subsubsection{Re-evaluating white-box membership inference attacks \cite{nasr_whitebox_privacy_SP19}}
\label{subsubsec:sp19}

We have shown that previous work may underestimate the target models' privacy risks, and the metric-based attacks with only black-box access can result in higher attack accuracy than NN based attacks for most models.
Recently Nasr et al. \cite{nasr_whitebox_privacy_SP19} demonstrated that a white-box membership inference adversary can perform stronger NN based attacks by using gradient with regard to model parameters.
Next, we evaluate whether the advantage of white-box attacks still exists by using our metric-based black-box benchmark attacks.

We follow Nasr et al. \cite{nasr_whitebox_privacy_SP19} to obtain classifiers on Purchase100, Texas100 and CIFAR100 datasets.
The Purchase100 classifier and the Texas100 classifier are same as undefended classifiers in Section \ref{subsubsec:ccs18}.
The CIFAR100 classifier is a publicly available pre-trained model,\footnote{\url{https://github.com/bearpaw/pytorch-classification}} with the DenseNet architecture \cite{densenet_huang_cvpr17}.
Table~\ref{tab:sp19_whitebox} lists all attack results.

From Table~\ref{tab:sp19_whitebox}, we can see that \textbf{compared to the black-box metric-based attacks, the improvement of white-box membership inference attacks is limited}. 
The attack accuracy of white-box membership inference adversary is only $0.5\%$ and $0.6\%$ higher than the attack accuracy achieved by our black-box benchmark attacks, on the Texas100 and the CIFAR100 classifiers.
The white-box attack on the Purchase100 classifier still has $5.8\%$ increase in attack accuracy compared to black-box attacks.
As a validation of our observations, we note that Shejwalkar and Houmansadr also report close membership inference attack accuracy between white-box attacks and black-box attacks in their recent work \cite{shejwalkar_mem_arxiv19}.

\section{Fine-Grained Analysis on Privacy Risks}\label{sec:fine-grained}

Prior work \cite{shokri_membership_SP17, song_privacyriks_CCS19, nasr_membership_defense_CCS18, nasr_whitebox_privacy_SP19, jia2019memguard_ccs19} focuses on an \emph{aggregate} evaluation of privacy risks by reporting overall attack accuracy or a precision-recall pair, which are averaged over all samples.
However, the target machine learning model's performance is usually varied across samples, which denotes the \emph{heterogeneity} of samples' privacy risks.
Therefore, a fine-grained privacy risk analysis of \emph{individual} samples is needed, with which we can understand the distribution of privacy risks over samples and identify which samples have high privacy risks.

In this section, we first define a metric called privacy risk score to quantitatively measure the privacy risks for each individual training member.
Then we use this metric to experimentally measure fine-grained privacy risks of target models. 
Overall, we argue that existing aggregate privacy analysis of ML models should be supplemented with our fine-grained privacy analysis for a thorough evaluation of privacy risks.

\subsection{Definition of privacy risk score}\label{subsec:risk_score}

For membership inference attacks, the privacy risk of a training member arises due to the distinguishability of its model prediction behavior with non-members.
This motivates our definition of the privacy risk score as following.
\begin{definition}\label{def:risk_score}
The privacy risk score of an input sample $\bfz = (\bfx, y)$ for the target machine learning model $F$ is defined as the posterior probability that it is from the training set $D_{\text{tr}}$ 
after observing the target model's behavior over that sample denoted as $\advobserv$, i.e.,
\end{definition}
\begin{equation}\label{eq:risk_score_definition}
r(\bfz)  = P(\bfz \in D_{\text{tr}} | \advobserv)
\end{equation}
Based on Bayes' theorem, we further compute the privacy risk score as following.

\begin{equation}\label{eq:risk_member}
\resizebox{1\hsize}{!}{$
\begin{aligned}
& r(\bfz)  = \frac{P(\bfz \in D_{\text{tr}}) \cdot P(\advobserv | \bfz \in D_{\text{tr}})}{P(\advobserv)} \\
& =  \frac{P(\bfz \in D_{\text{tr}})  \cdot P(\advobserv | \bfz \in D_{\text{tr}})}{P(\bfz \in D_{\text{tr}})  \cdot P(\advobserv | \bfz \in D_{\text{tr}}) + P(\bfz \in D_{\text{te}}) \cdot P(\advobserv| \bfz \in D_{\text{te}})} ,
\end{aligned}
$
}
\end{equation}
where $D_{\text{te}}$ stands for the test set. The observation $\advobserv$ depends on the adversary's access to the target model: in the black-box membership inference attack \cite{shokri_membership_SP17}, it is the model's final output, i.e., $\advobserv = F(\bfx)$; in the white-box membership inference attacks \cite{nasr_whitebox_privacy_SP19}, it also includes the model's intermediate layers' outputs and gradient information at all layers.
Our proposed benchmark attacks only need black-box access to the target model, and most existing attack methods \cite{shokri_membership_SP17, yeom_privacy_CSF18, song_privacyriks_CCS19} work in the black-box manner. Therefore, we focus on the black-box scenario for the computation of the privacy risk score in this paper and leave the discussion on white-box scenario as future work. In the black-box attack scenario, the privacy risk score can be expressed as
\begin{equation}\label{eq:risk_member_2}
\resizebox{1\hsize}{!}{$
\begin{aligned}
& r(\bfz)
 = \frac{P(\bfz \in D_{\text{tr}})  \cdot P(F(\bfx) | \bfz \in D_{\text{tr}})}{P(\bfz \in D_{\text{tr}})  \cdot P(F(\bfx) | \bfz \in D_{\text{tr}}) + P(\bfz \in D_{\text{te}}) \cdot P(F(\bfx) | \bfz \in D_{\text{te}})} 
\end{aligned}
$
}
\end{equation}

From Equation \eqref{eq:risk_member_2}, we can see that the risk score depends on both prior probabilities $P(\bfz \in D_{\text{tr}})$, $P(\bfz \in D_{\text{te}})$ and conditional distributions $P(F(\bfx)|\bfz \in D_{\text{tr}})$, $P(F(\bfx)|\bfz \in D_{\text{te}})$.
For the prior probabilities, we follow previous work \cite{shokri_membership_SP17, yeom_privacy_CSF18} to assume that an example is sampled from either training set or test set with an equal $0.5$ probability, where the uncertainty of membership inference attacks is maximized. Note that the privacy risk score is naturally applicable to any prior probability scenario, and we present the results with different prior probabilities in Appendix \ref{appendix:different_priors}. With the equal probability assumption, we have
\begin{equation}\label{eq:risk_member_3}
r(\bfz)
 = \frac{  P(F(\bfx) | \bfz \in D_{\text{tr}})}{  P(F(\bfx) | \bfz \in D_{\text{tr}}) +  P(F(\bfx) | \bfz \in D_{\text{te}})} 
\end{equation}

For the conditional distributions $P(F(\bfx)|\bfz \in D_{\text{tr}})$, $P(F(\bfx)|\bfz \in D_{\text{te}})$, we empirically measure these values using shadow-training technique: (1) train a shadow model to simulate the behavior of the target model; (2) obtain the shadow model's prediction outputs on shadow training and shadow test data; (3) empirically compute the conditional distributions on shadow training and shadow test data.
Furthermore, as the class-dependent thresholding technique is shown to improve the attack success in Table~\ref{tab:ccs18_threhsold_setting}, we compute the distribution of model prediction over training data $P(F(\bfx) | \bfz \in D_{\text{tr}})$ in a class-dependent manner ($P(F(\bfx)|\bfz \in D_{\text{te}})$ is computed in the same way). 
\begin{equation}\label{eq:distribution_1}
\resizebox{1\hsize}{!}{$
P(F(\bfx) | \bfz \in D_{\text{tr}}) =  \left\{
\begin{aligned}
P(F(\bfx) | \bfz \in D_{\text{tr}}, y=y_{0}), & \quad \text{when} \quad y=y_{0} \\
P(F(\bfx) | \bfz \in D_{\text{tr}}, y=y_{1}), & \quad \text{when} \quad y=y_{1}\\
& \vdots\\
P(F(\bfx) | \bfz \in D_{\text{tr}}, y=y_{n}), & \quad \text{when} \quad y=y_{n}
\end{aligned}
\right.
$}
\end{equation}

Since we empirically measure the conditional distributions using the shadow model's predictions over shadow data, the quality of measured distributions highly depends on the shadow model's similarity to the target model and the size of shadow data. 
On the one hand, the size of shadow data is usually limited. Especially in our analysis where the distribution is computed in a class-dependent manner, for each class label $y_n$, we may not have enough samples\footnote{In our experiments, on average we have 197 samples per class for Purchase100 dataset; 100 samples per class for Texas100 dataset; 33 samples per class for Location30 dataset; and 500 per class for CIFAR100 dataset.} to adequately estimate the multi-dimension distribution $P(F(\bfx) | \bfz \in D_{\text{tr}}, y=y_{n})$.
On the other hand, in Section \ref{subsec:MIA_results} we show that by only using the one-dimension prediction metric such as confidence and modified entropy, our proposed benchmark attacks in fact achieve comparable or even better success that NN-based attacks which leverage the whole prediction vector as features.
Thus, we propose to further approximate the multi-dimension distribution in Equation \eqref{eq:distribution_1} with the distribution of modified prediction entropy, since using modified entropy usually results in highest attack accuracy among all benchmark attacks.\footnote{In most cases, both modified entropy based attack and confidence based attack give best attack performance. However, for undefended Location30 and Texas100 classifiers in Table~\ref{tab:ccs19_defense}, the modified entropy based attack achieves significantly higher attack accuracy.}

\begin{equation}\label{eq:distribution_2}
\resizebox{1\hsize}{!}{$
P(F(\bfx) | \bfz \in D_{\text{tr}}) \approx  \left\{
\begin{aligned}
P(\mentropy(F(\bfx), y) | \bfz \in D_{\text{tr}}, y=y_{0}), & \quad \text{when} \quad y=y_{0} \\
P(\mentropy(F(\bfx), y) | \bfz \in D_{\text{tr}}, y=y_{1}), & \quad \text{when} \quad y=y_{1}\\
& \vdots\\
P(\mentropy(F(\bfx), y) | \bfz \in D_{\text{tr}}, y=y_{n}), & \quad \text{when} \quad y=y_{n}
\end{aligned}
\right.
$}
\end{equation}
We also approximate $P(F(\bfx) | \bfz \in D_{\text{te}})$ in the same way.
By plugging Equation \eqref{eq:distribution_2} into Equation \eqref{eq:risk_member_3}, we can get the privacy risk score for a certain sample.

\subsection{Experiment results}

\begin{figure*}[!ht]
\centering
    \begin{subfigure}[t]{\linewidth}
		\centering
		\includegraphics[width=\linewidth]{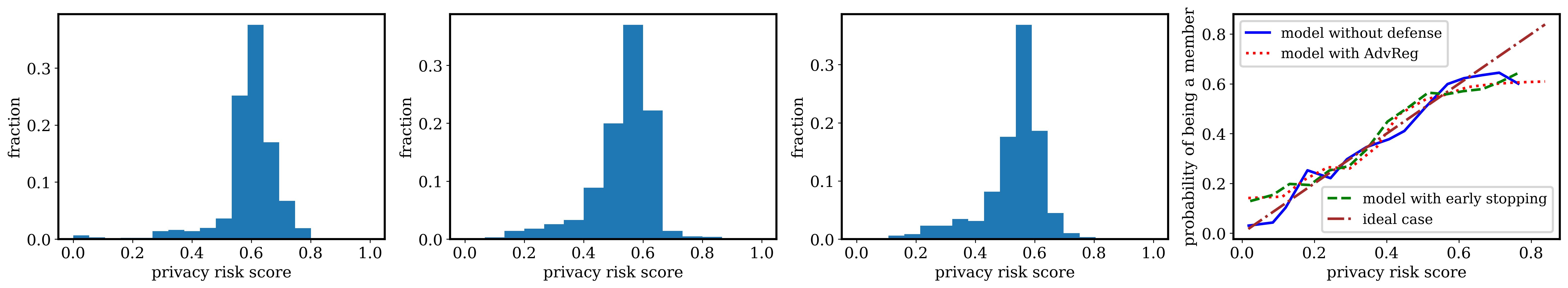}
		\caption{
		The first three figures present the distributions of training samples' privacy risk scores on Purchase100 classifiers without defense, with AdvReg \cite{nasr_membership_defense_CCS18}, and with early stoppping.
		The last figure shows that the privacy risk score can well represent the real probability of being a member, with root mean square error (RMSE) of 0.05, 0.09, and 0.06.
		}\label{fig:purchase_valid}
	\end{subfigure}\hfill
	\begin{subfigure}[t]{\linewidth}
		\centering
		\includegraphics[width=\linewidth]{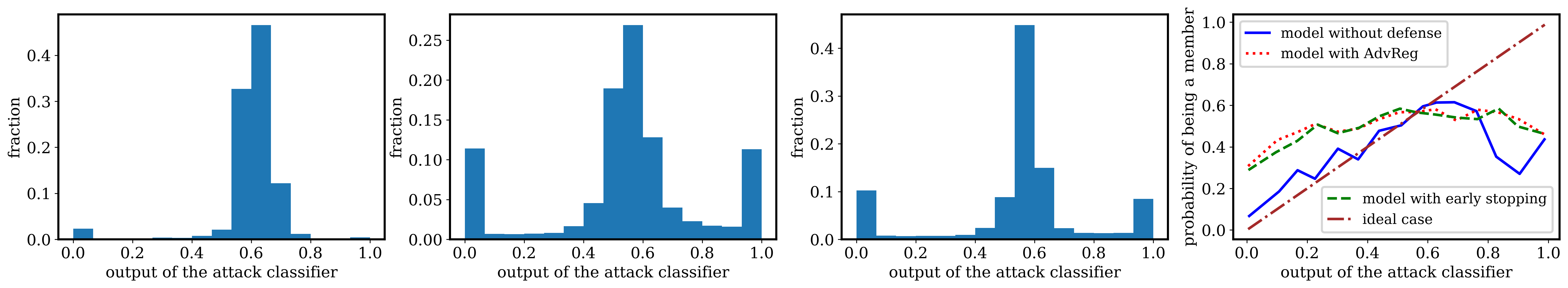}
		\caption{
		The first three figures present the distributions of NN attack classifier's outputs over training samples on Purchase100 classifiers without defense, with AdvReg \cite{nasr_membership_defense_CCS18}, and with early stoppping.
		The last figure shows that the NN classifier's output fails to represent the real probability of being a member, with RMSE values of 0.26, 0.26, and 0.25.
		}\label{fig:NN_purchase_valid}
	\end{subfigure}\hfill
\vspace{-0.5em}\caption{
Estimate the probability of being a member with our proposed privacy risk score (Figure~\ref{fig:purchase_valid}), and with the NN attack classifier's output (Figure~\ref{fig:NN_purchase_valid}).
}
\label{fig:risk_score_valid_purchase}
\end{figure*}

\label{subsec:risk_score_experiments}

In our experiments, we first validate that our proposed privacy risk score really captures the \emph{probability} of being a member.
Next, we compare the distributions of training samples' privacy risk scores for target models without defense and with defenses \cite{nasr_membership_defense_CCS18, jia2019memguard_ccs19}.
We then demonstrate how to use privacy risk scores to perform membership inference attacks with high confidence.
Finally, we perform an in-depth investigation of individual samples' privacy risk scores by correlating them with model sensitivity, generalization errors, and feature embeddings.
To have enough diversity of data and models, and to further evaluate defense methods, we perform experiments on 3 Purchase100 classifiers (without defense, with AdvReg \cite{nasr_membership_defense_CCS18}, and with early stopping) and 2 Texas100 classifiers (without defense, and with MemGuard \cite{jia2019memguard_ccs19}).
Both Purchase100 classifiers and Texas classifiers use fully connected neural networks with 4 hidden layers, and the numbers of neurons for hidden layers are 1024, 512, 256, and 128, respectively. Purchase100 classifiers use Tanh as the activation function \cite{nasr_membership_defense_CCS18}, and Texas100 classifiers use ReLU as the activation function \cite{jia2019memguard_ccs19}.

\subsubsection{Validation of privacy risk score}\label{subsubsec:validation}

Before presenting the detailed results for privacy risk score, we first validate its effectiveness here.
For the target machine learning model, we first compute the privacy risk scores following the method in Section \ref{subsec:risk_score} for all training and test samples.
Next we divide the entire range of privacy risk scores into multiple bins, and count the number of training points ($n_{\text{tr}}$) and the number of test points ($n_{\text{te}}$) in each bin. Then we compute the fraction of training points ($\frac{n_{\text{tr}}}{n_{\text{tr}}+n_{\text{te}}}$) in each bin, which indicates the real likelihood of a sample being a member (y axis of the last column in Figure \ref{fig:purchase_valid}).
If the privacy risk score truly corresponds to the probability that a sample is from a target model’s training set, then we expect the actual values of privacy risk scores and fraction of training points in each bin to closely track with each other.

As a baseline to compare with, we also consider using NN based attacks to estimate privacy risks of individual samples.
Prior papers suggest using the attack classifier's prediction to measure the input's privacy risk \cite{nasr_membership_defense_CCS18, jia2019memguard_ccs19}.
The attack classifier has only one output, which is within [0, 1] and can serve as a proxy to estimate the probability of being a member.
Following same steps as above, we compute the real probability of being a member and the average outputs of the attack classifier.
Specifically, we follow Nasr et al. \cite{nasr_membership_defense_CCS18} to train the attack classifier by using the target model's predictions and one-hot encoded input labels as features.

Figure~\ref{fig:risk_score_valid_purchase} shows the distribution of training samples' privacy risk scores (top row) and attack classifier's outputs on training data (bottom row) for Purchase100 classifiers without defense, with AdvReg \cite{nasr_membership_defense_CCS18}, and with early stopping.
We also compare the privacy risk score and attack classifier's output with the real probability of being a member, as shown in the last column of Figure~\ref{fig:risk_score_valid_purchase} where the ideal case is used to check the effectiveness of metrics.
We can see that \textbf{our proposed privacy risk score closely aligns with the actual probability of being a member}: the privacy risk score curves for all three models are quite close to the line of the ideal case.
On the other hand, the attack classifiers' outputs fail to capture the membership probability.
This is because the NN classifiers are trained to minimize the loss, i.e., the output of a member should be close to 1 while the output of a non-member should be close to 0.
With this training goal, the obtained attack classifiers failed to capture the privacy risks for individual samples.
We also quantitatively measure the root-mean-square error (RMSE) between estimated probability of member
and real probability of member.
On the three Purchase100 classifiers, the RMSE values of our privacy risk score are 0.05, 0.09, and 0.06; in contrast, the RMSE values of NN classifier's outputs are 0.26, 0.26, 0.25, respectively.
We observe similar results on the undefended Texas100 classifier and the defended classifier by MemGuard \cite{jia2019memguard_ccs19}, with details in Appendix \ref{appendix:texas_score_valid}.

\begin{figure*}[!ht]
	\centering
	\begin{subfigure}[t]{0.31\linewidth}
		\raggedleft
		\includegraphics[width=\linewidth]{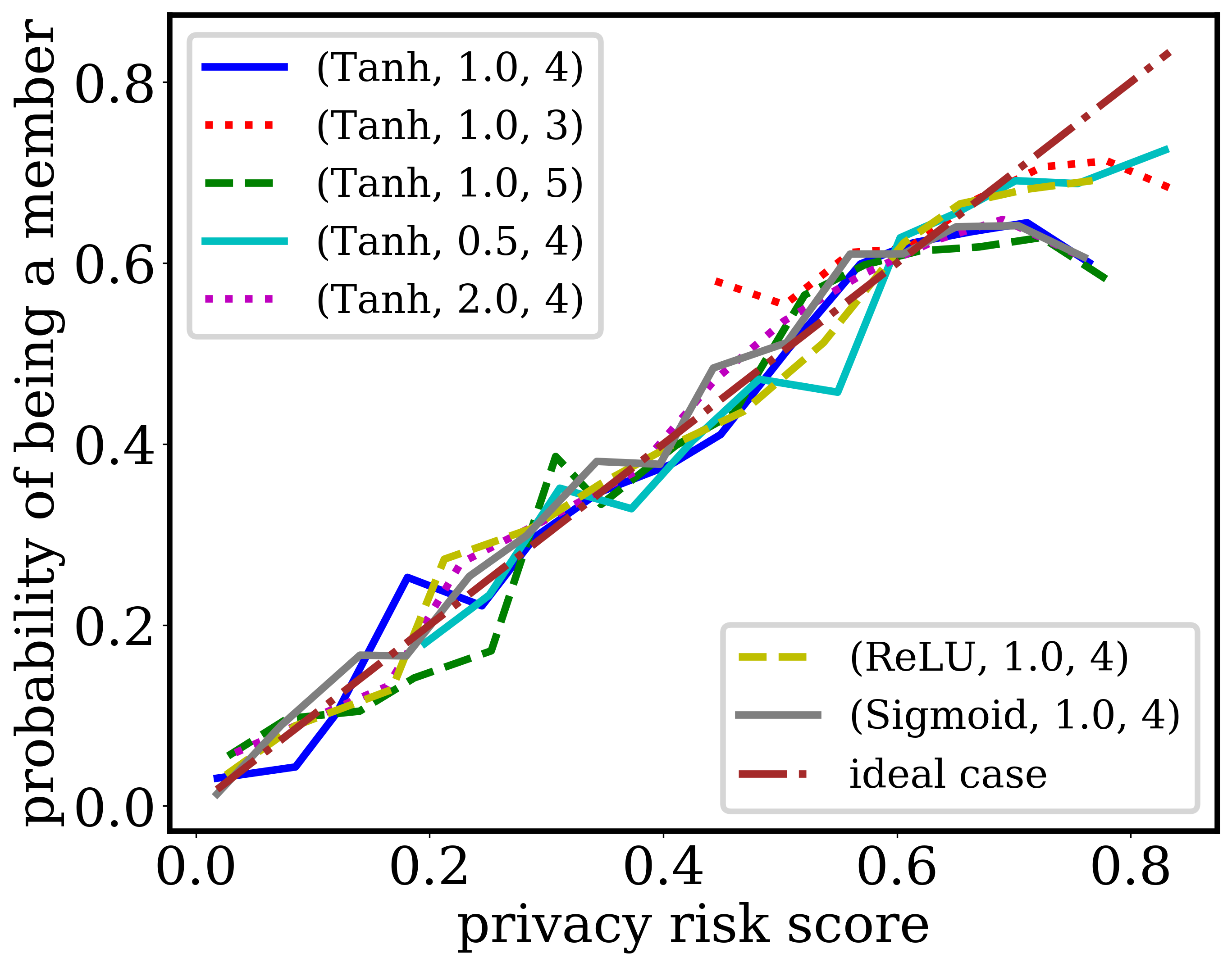}
	\end{subfigure}\hfill
	\begin{subfigure}[t]{0.31\linewidth}
		\centering
		\includegraphics[width=\linewidth]{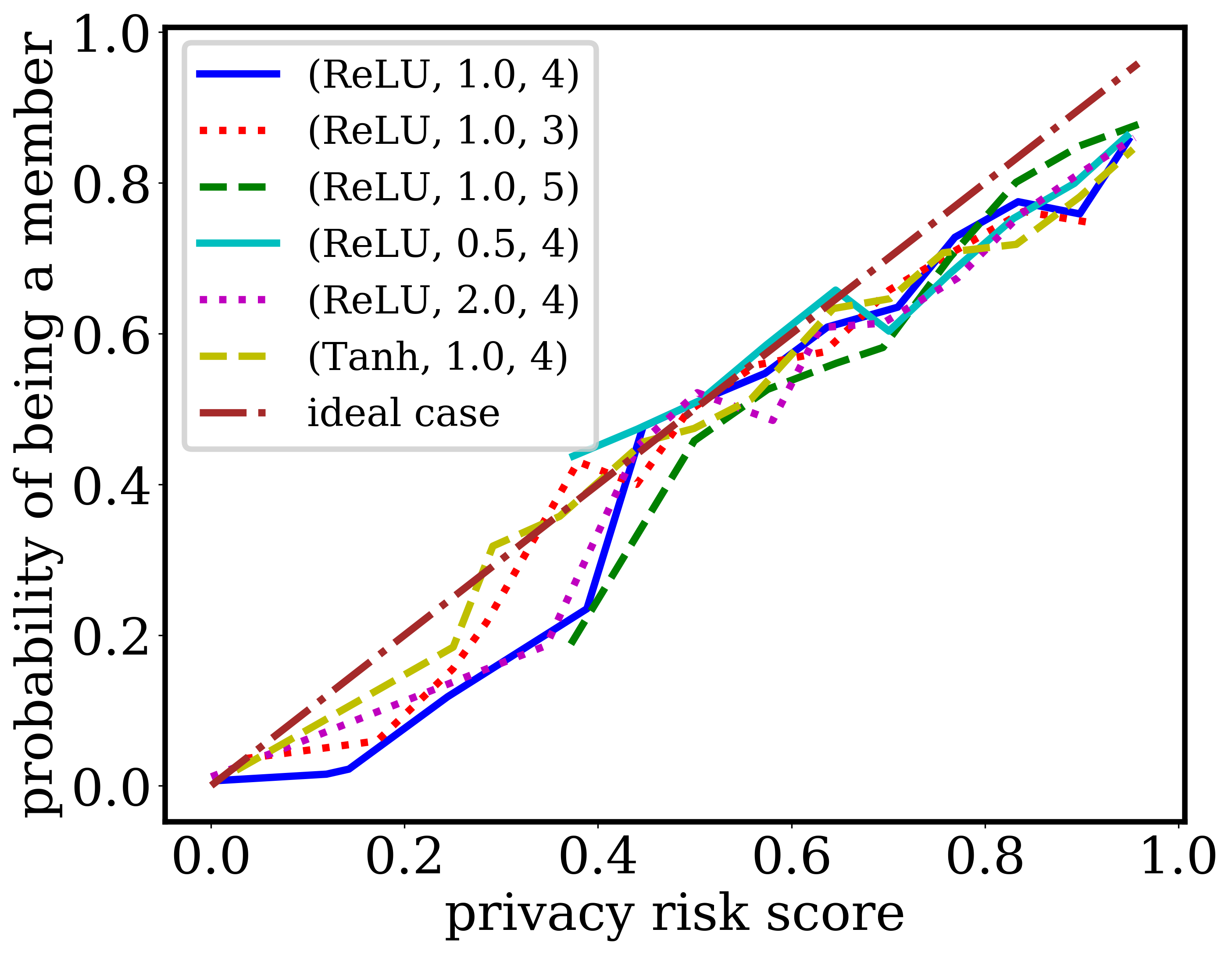}
	\end{subfigure}\hfill
	\begin{subfigure}[t]{0.31\linewidth}
		\raggedright
		\includegraphics[width=\linewidth]{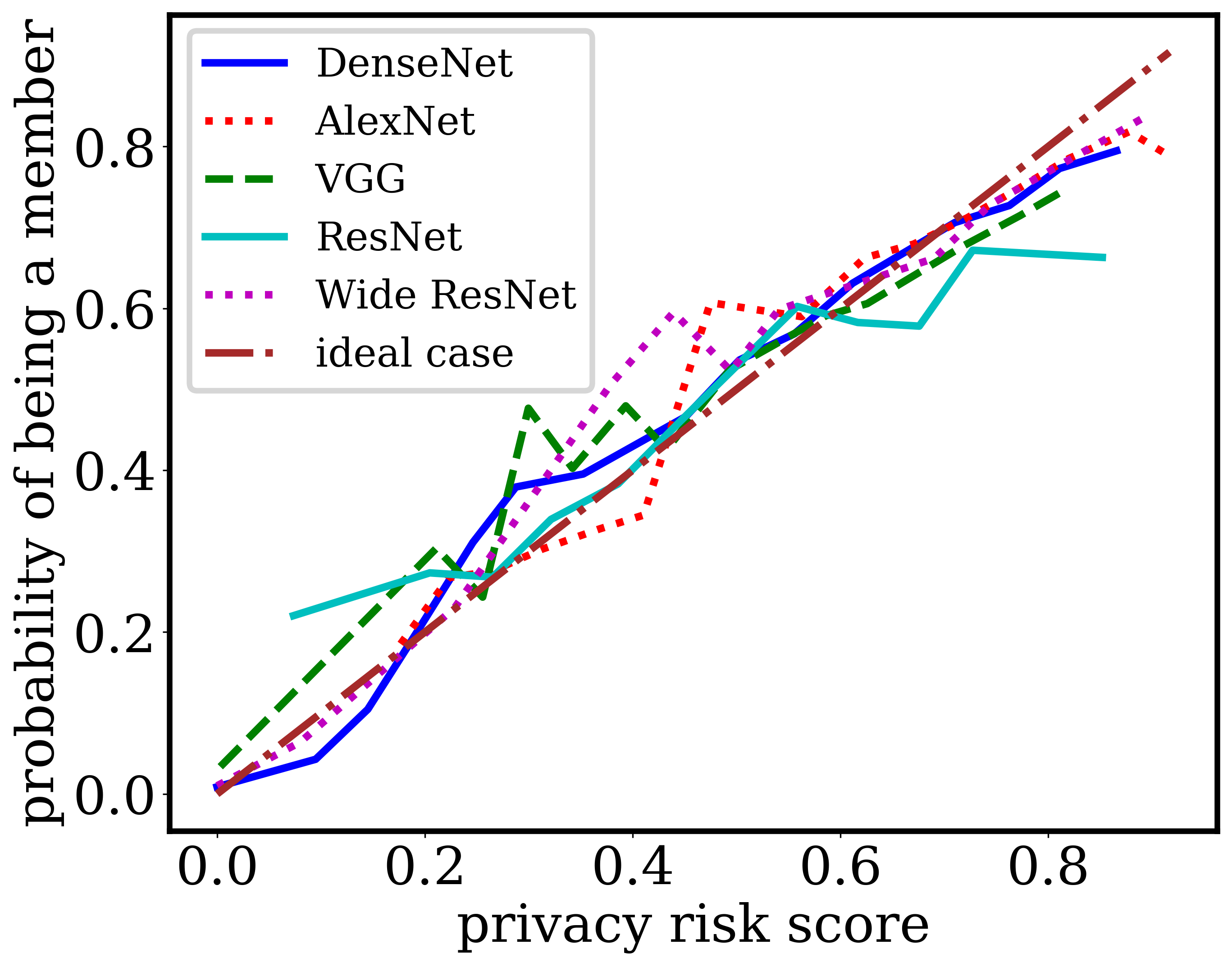}
	\end{subfigure}\hfill
	\vspace{-0.5em}\caption{Validation of privacy risk score with different model architectures on 
	(undefended) Purchase100 (left), Texas100 (middle), and CIFAR100 (right) classifiers. 
	For Purchase100 and Texas100 classifiers, the legend is expressed as (activation function, width, depth). 
	The RSME values between privacy risk score (x-axis) and probability of being a member (y-axis) for all lines are smaller than 0.09.
	}
	\label{fig:more_validation}
\end{figure*}

We also validate the effectiveness of privacy risk score across varied model architectures. 
For Purchase100 and Texas100 classifiers, we test two additional neural network depths by deleting the last hidden layer (depth=3) or adding one more hidden layer with 2048 neurons (depth=5); we test two additional neural network widths by halving the numbers of hidden neurons (width=0.5) or doubling the numbers of hidden neurons (width=2.0); we also test ReLU, Tanh, or Sigmoid as the activation functions.
For CIFAR100 classifiers, besides DenseNet \cite{densenet_huang_cvpr17}, we test other popular convolutional neural network architectures, including AlexNet \cite{krizhevsky_imagenet_NIPS12}, VGG \cite{simonyan_DNN_arxiv14}, ResNet \cite{he_ResNet_CVPR16}, and Wide ResNet \cite{Zagoruyko_WRN_BMVC16}.
As show in Figure \ref{fig:more_validation}, our proposed privacy risk score metric indeed well represents the likelihood of a sample being in the training set under different model architectures.
On the Texas100 dataset, the classifier fails to learn meaningful features using the Sigmoid activation function, achieving an  accuracy of only 4\%, and is thus omitted from the figure. 
We provide validation results with defended classifiers in Appendix \ref{appendix:validation_varied_architecture}.

\subsubsection{Heterogeneity of members' privacy risk scores}
After validating the effectiveness of the privacy risk score metric, 
we show the heterogeneity of training samples' privacy risks by plotting the cumulative distribution of their privacy risk scores.
We also investigate the performance of membership inference defense methods \cite{nasr_membership_defense_CCS18, jia2019memguard_ccs19} with comparison between defended and undefended classifiers.

\begin{figure}[!ht]
	\centering
	\includegraphics[width=\linewidth]{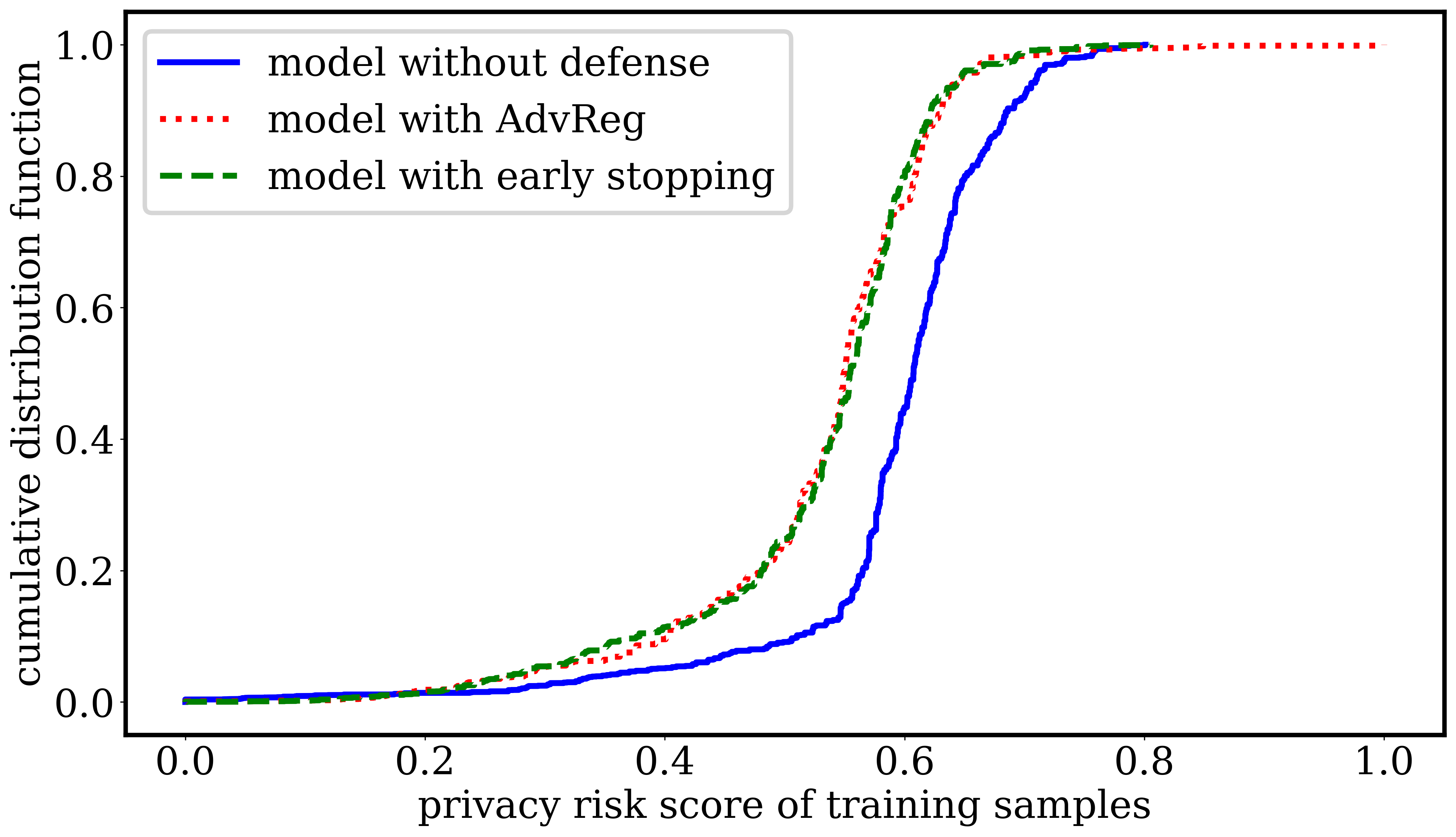}
	\caption{The cumulative distribution of privacy risk scores for Purchase100 classifiers in Nasr et al. \cite{nasr_membership_defense_CCS18}.
	}
	\label{fig:score_cdf1}
\end{figure}

Figure~\ref{fig:score_cdf1} presents the cumulative distributions of training points' privacy risk scores for Purchase100 classifiers. 
We can see that, compared with the undefended classifier, \textbf{the defended classifier with adversarial regularization \cite{nasr_membership_defense_CCS18} has smaller privacy risk scores on average}.
However, we can also see that \textbf{the defended classifier has a small portion of training data with higher privacy risk scores than the undefended model}: the undefended model has all members' privacy risk scores under 0.8, in contrast, the defended model has several training points with privacy risk scores higher than 0.8.
Furthermore, \textbf{the classifier with early stopping has a similar risk score distribution as the defended classifier}.

\begin{figure}[!ht]
	\centering
	\includegraphics[width=\linewidth]{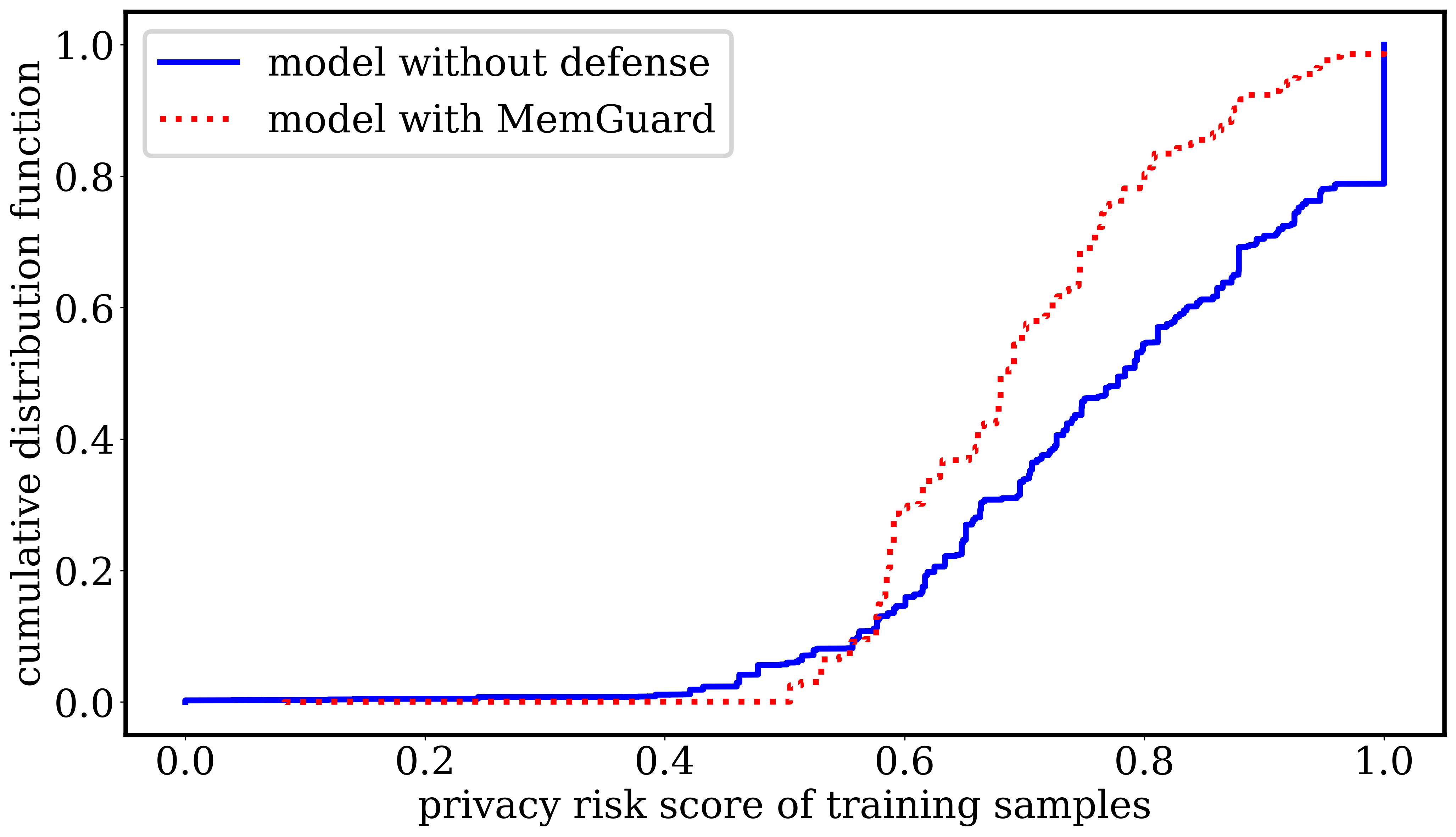}
	\caption{The cumulative distribution of privacy risk scores for Texas100 classifiers in Jia et al. \cite{jia2019memguard_ccs19}.
	}
	\label{fig:score_cdf2}\vspace{-2mm}
\end{figure}
Figure~\ref{fig:score_cdf2} shows the cumulative distribution of training data' privacy risk scores for Texas100 classifiers. 
We can see that \textbf{the defense method indeed decreases training samples' privacy risk scores}.
However, \textbf{the defended classifier is still quite vulnerable}: $70\%$ training samples have privacy risk scores higher than 0.6.

\subsubsection{Usage of privacy risk score}\label{subsubsec:score_usage}
From our definition and verification results in Section~\ref{subsubsec:validation}, we know that 
privacy risk score of a data point indicates its probability of being a member.
Instead of pursuing high \emph{average} attack accuracy, now the adversary can identify which samples have high privacy risks and perform \emph{attacks with high confidence}: a sample is inferred as a member if and only if its privacy risk score is above a certain probability threshold.

\begin{table*}[!ht]
\caption{Membership inference attacks by setting different threshold values on privacy risk scores.
For each threshold value, we report the (precision, recall) pair of membership inference attacks.
}
\centering
\renewcommand\arraystretch{1.0}
\fontsize{8.1pt}{8.1pt}\selectfont
\begin{tabular}{cccccccc}
\toprule[1.5pt]
\multirow{2}{*}{\textbf{dataset}} & \textbf{defense} & \multicolumn{6}{c}{\textbf{\normalsize{threshold values on privacy risk scores}}} \\
& \textbf{method} & 1 & 0.9 & 0.8 & 0.7 & 0.6 & 0.5 \\
\midrule[0.75pt]
\multirow{4}{*}{Texas100} & \multirow{2}{*}{no defense} & \multirow{2}{*}{(85.4\%, 21.2\%)} & \multirow{2}{*}{(83.4\%, 29.1\%)} & \multirow{2}{*}{(81.2\%, 45.3\%)} &  \multirow{2}{*}{(77.0\%, 66.1\%)} & \multirow{2}{*}{(72.8\%, 85.4\%)} & \multirow{2}{*}{(70.6\%, 94.3\%)}\\
& & & & & & &  \\
& \multirow{2}{*}{MemGuard \cite{jia2019memguard_ccs19} } & \multirow{2}{*}{(88.2\%, 1.4\%)} & \multirow{2}{*}{(84.5\%, 7.6\%)} & \multirow{2}{*}{(82.6\%, 18.7\%)} &  \multirow{2}{*}{(77.0\%, 43.7\%)} & \multirow{2}{*}{(71.3\%, 70.5\%)} & \multirow{2}{*}{(66.0\%, 99.9\%)}\\
& & & & & & &  \\
\midrule[0.75pt]
\multirow{6}{*}{Purchase100} & \multirow{2}{*}{no defense} & \multirow{2}{*}{(N.A., 0\%)} & \multirow{2}{*}{(N.A., 0\%)} & \multirow{2}{*}{(N.A., 0\%)} &  \multirow{2}{*}{(63.4\%, 7.8\%)} & \multirow{2}{*}{(62.6\%, 55.1\%)} & \multirow{2}{*}{(61.6\%, 90.9\%)}\\
& & & & & & &  \\
& \multirow{2}{*}{early stopping} & \multirow{2}{*}{(N.A., 0\%)} & \multirow{2}{*}{(N.A., 0\%)} & \multirow{2}{*}{(N.A., 0\%)} &  \multirow{2}{*}{(60.4\%, 1.3\%)} & \multirow{2}{*}{(57.1\%, 20.2\%)} & \multirow{2}{*}{(56.6\%, 75.2\%)}\\
& & & & & & &  \\
& \multirow{2}{*}{AdvReg \cite{nasr_membership_defense_CCS18} } & \multirow{2}{*}{(83.3\%, 0.2\%)} & \multirow{2}{*}{(83.3\%, 0.2\%)} & \multirow{2}{*}{(65.9\%, 0.5\%)} &  \multirow{2}{*}{(63.9\%, 1.7\%)} & \multirow{2}{*}{(58.9\%, 24.6\%)} & \multirow{2}{*}{(56.5\%, 76.3\%)}\\
& & & & & & &  \\
\bottomrule[1.5pt]
\end{tabular}
\vspace{-0.5em}
\label{tab:attacks_with_privacy_threshold}
\end{table*}

We show the attack results with precision and recall values in Table~\ref{tab:attacks_with_privacy_threshold} for target classifiers with varying threshold values on privacy risk scores.
From Table~\ref{tab:attacks_with_privacy_threshold}, we can see that with larger threshold values on privacy risk scores, the adversary indeed has higher precision values for membership inference attacks.
For MemGuard \cite{jia2019memguard_ccs19}, when setting the same threshold value on privacy risk scores, both undefended and defended Texas100 classifiers have similar attack precision, but the defended classifier has a smaller recall value.
However, the defended Texas100 classifier still has severe privacy risks: $70.5\%$ training members can be inferred correctly with the precision of $71.3\%$, and $1.4\%$ training members can be inferred correctly with the precision of $88.2\%$. 
Similarly, while  adversarial regularization \cite{nasr_membership_defense_CCS18} can lower the average privacy risks, it increases the privacy risks for certain members: on the defended Purchase100 classifier, $0.2\%$ training members can be inferred correctly with the precision of $83.3\%$. We urge designers of defense mechanisms to thus account for the full distribution of privacy risks in their analysis.

\subsubsection{Impact of model properties on privacy risk score }\label{subsubsec:score_properties}
We perform an in-depth investigation of privacy risk score by exploring its correlations with certain model properties, including sensitivity, generalization error, and feature embedding.
We use the undefended Texas100 classifier from Jia et al. \cite{jia2019memguard_ccs19} for the following experiments.

\begin{figure}[!ht]
	\centering
	\includegraphics[width=\linewidth]{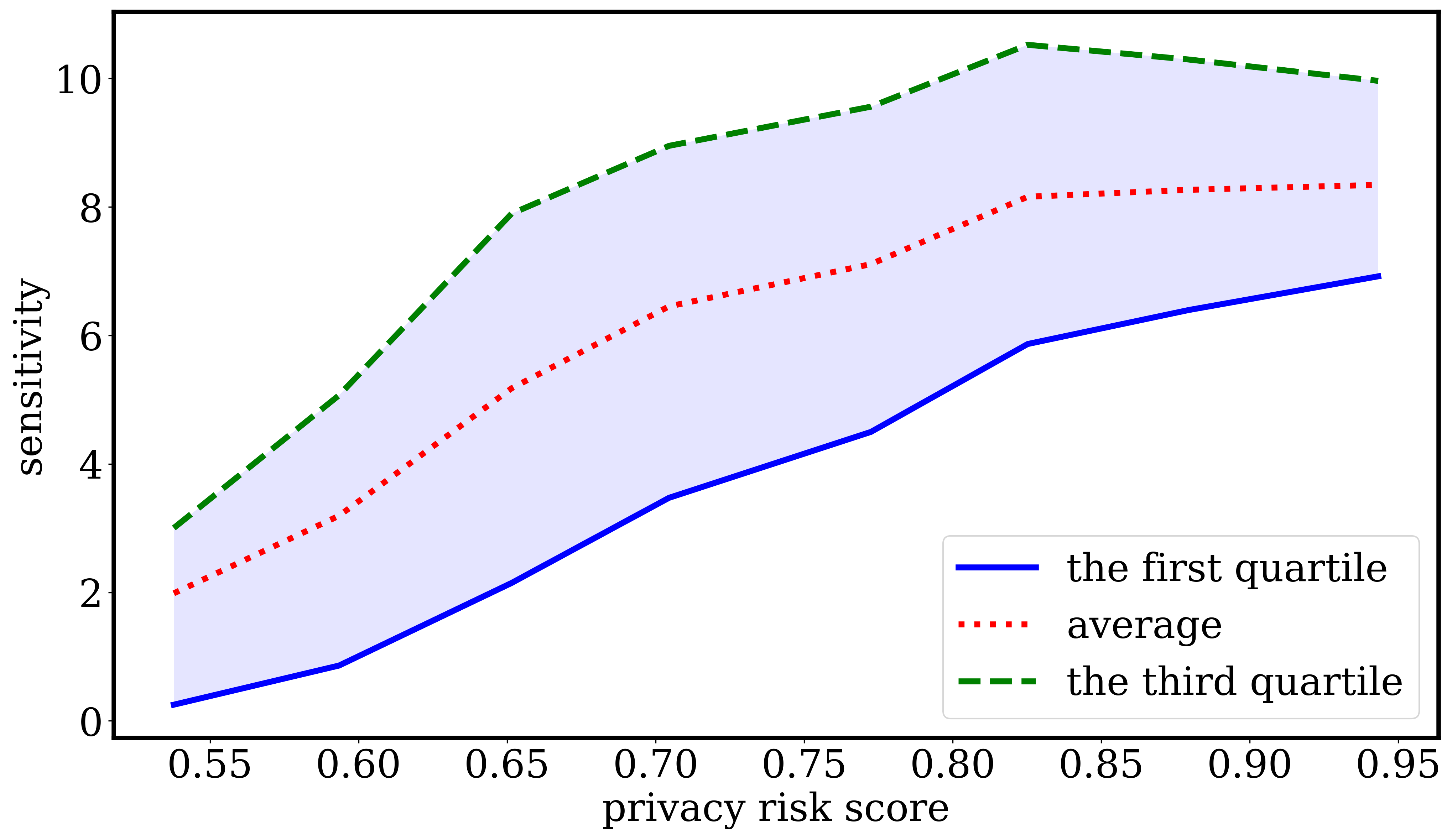}
	\caption{The relation between privacy risk score and model sensitivity.
	}
	\label{fig:sen_risk2}\vspace{-2mm}
\end{figure}

\noindent \textbf{{Privacy risk score with sensitivity}}
We first study the relationship between privacy risk scores and model sensitivity with regard to training samples.
The sensitivity is defined as the influence of one training sample on the target model by computing the difference after removing that sample.
Since the privacy risk score 
is obtained with the measured distributions of modified prediction entropy (Equation \eqref{eq:distribution_2}), we compute the model's sensitivity regard to a training point $\bfz = (\bfx, y)$ as the logarithm of $\frac{\mentropy(\widetilde{F}_{\bfz}(\bfx), y) }{\mentropy(F(\bfx), y)}$, where $\widetilde{F}_{\bfz}$ means the retrained classifier after removing $\bfz$ from the training set.

Figure~\ref{fig:sen_risk2} shows the relation between privacy risk scores and the model sensitivity.
For each privacy risk score, we show the first quartile, the average, and the third quartile of model sensitivities with regard to training data.
We can see that, \textbf{samples with higher privacy risk scores are likely to have a larger influence on the target model}.

\noindent \textbf{Privacy risk score with generalization error}
We observe that training samples with high risk scores are typically concentrated in a few class labels.
Therefore, we further compare privacy risk scores among different class labels in this section.

\begin{figure}[!ht]
	\centering
	\includegraphics[width=\linewidth]{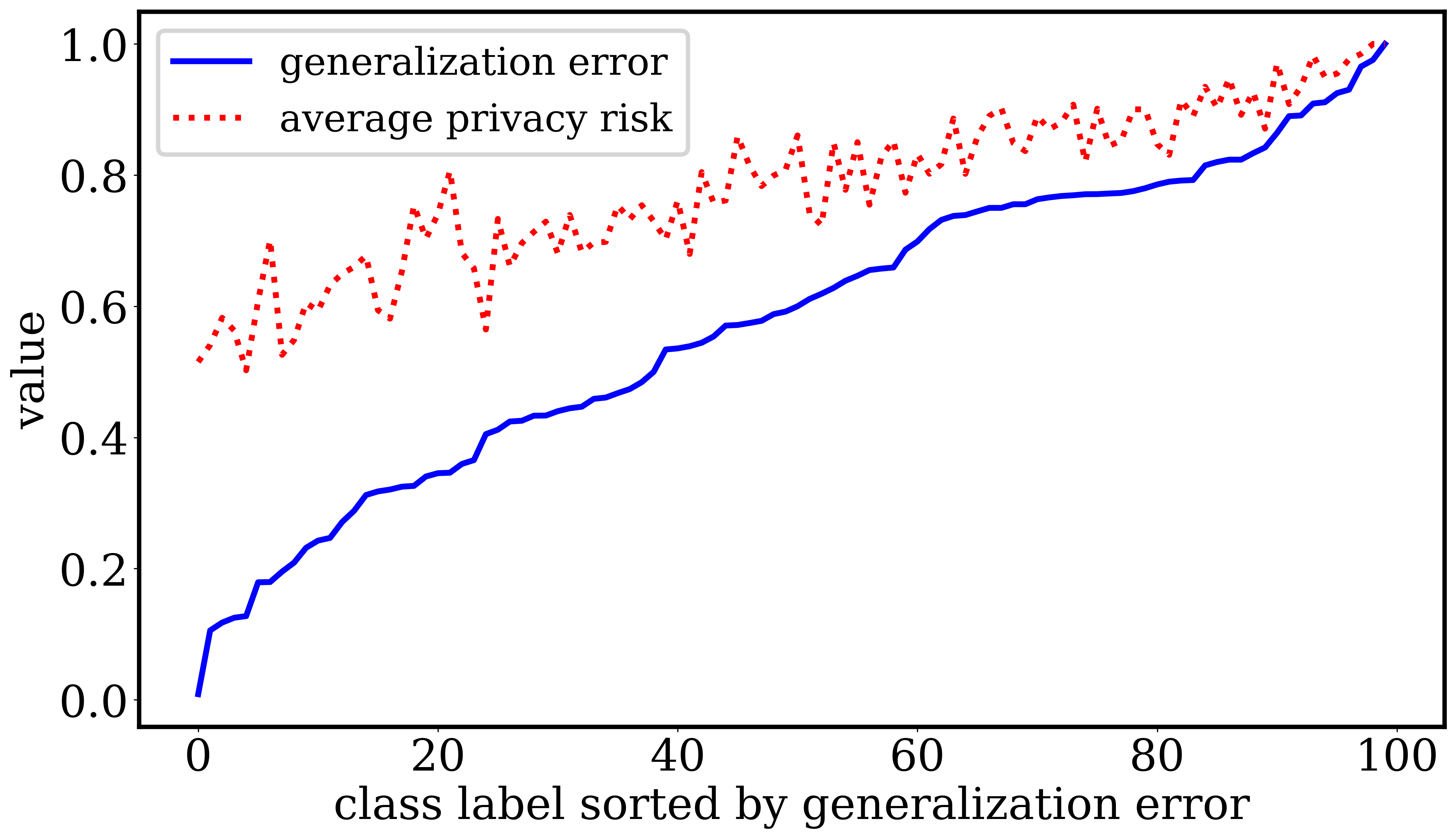}
	\caption{Average privacy risk score vs generalization error per class with a strong Pearson correlation coefficient of 0.94.
	}
	\label{fig:err_risk}\vspace{-2mm}
\end{figure}

Besides the privacy risk scores, we also record the generalization errors for different class labels. 
Figure~\ref{fig:err_risk} shows the average privacy risk scores and generalization errors for all 100 classes, where we sort the class labels based on their generalization errors.
We can see that \textbf{the class labels with high generalization errors tend to have higher privacy risk scores}, which is as expected since the generalization error has a large influence on the success of membership inference attacks \cite{shokri_membership_SP17}.
The Pearson correlation coefficient between average privacy risk scores and generalization errors is as high as 0.94.

\begin{figure*}[!ht]
\centering
    \begin{subfigure}[t]{0.44\linewidth}
		\raggedleft
		\includegraphics[width=\linewidth]{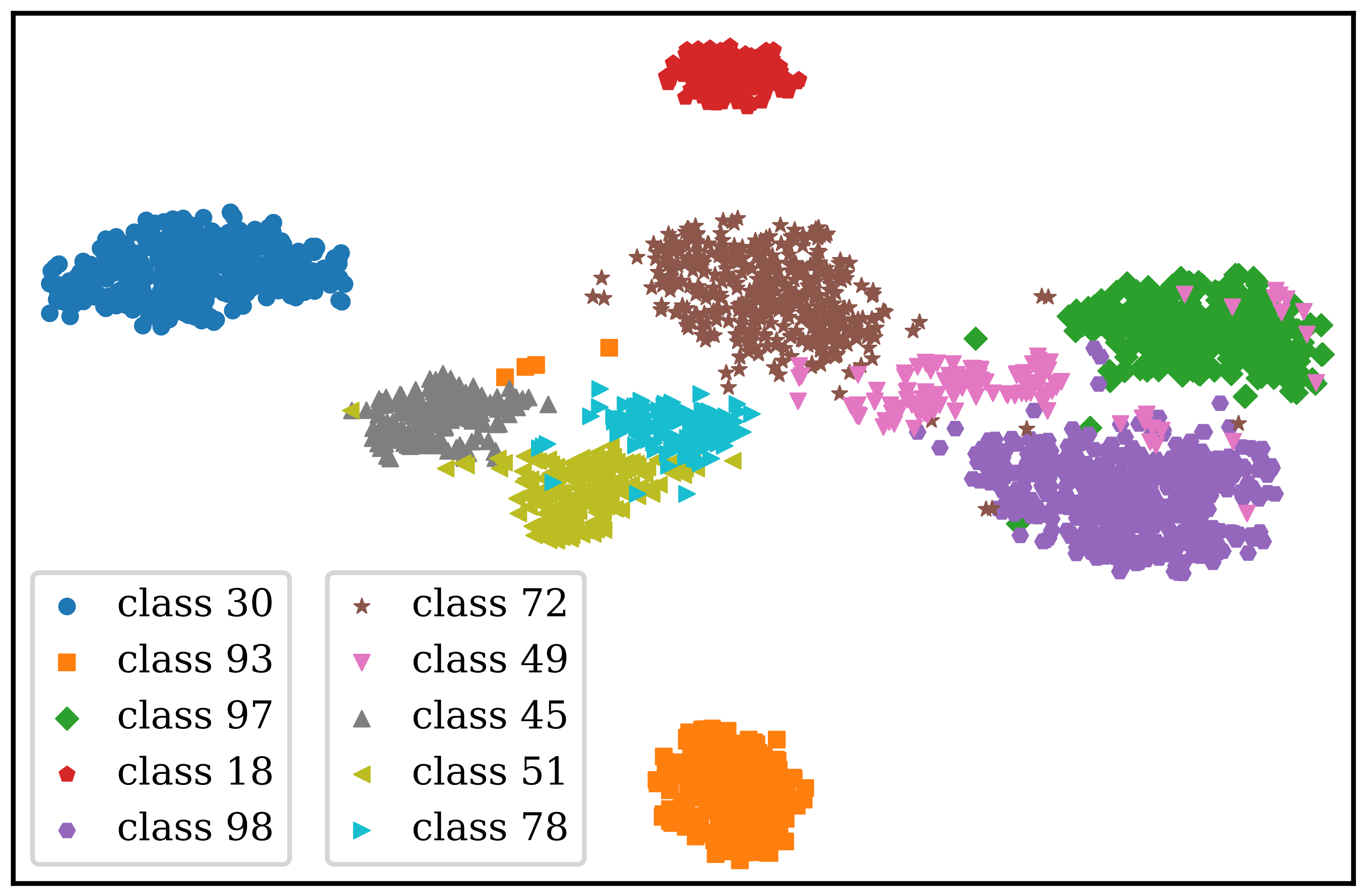}
		\caption{
		t-SNE plot for training samples in 10 class labels.
		}
		\label{fig:tsne_train}
	\end{subfigure}\hfill
	\begin{subfigure}[t]{0.44\linewidth}
		\raggedright
		\includegraphics[width=\linewidth]{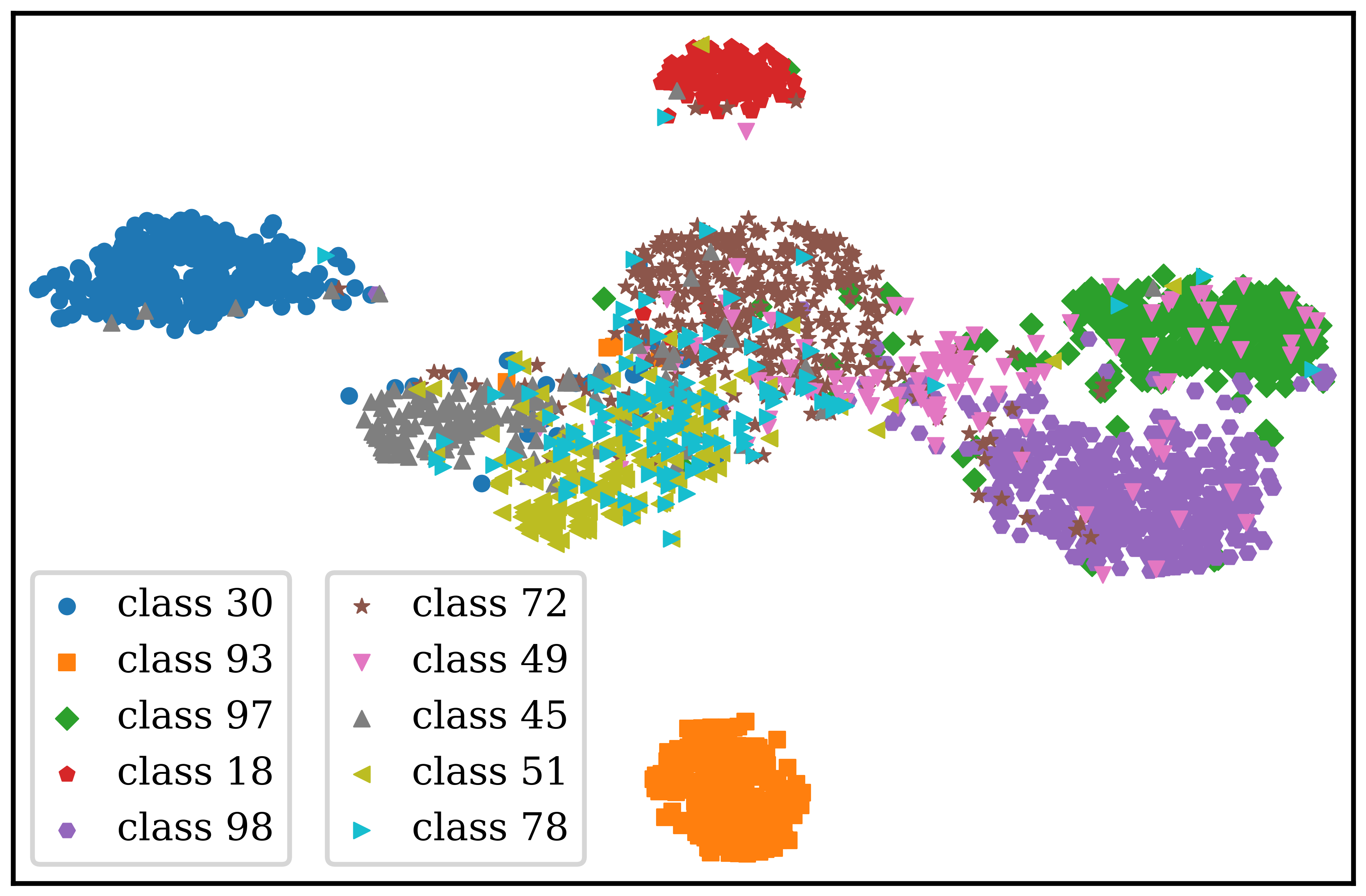}
		\caption{
		t-SNE plot for test samples in 10 class labels
		}
		\label{fig:tsne_test}
	\end{subfigure}
	\vspace{-0.5em}
\caption{
By using t-SNE \cite{tsne_maaten_JMLR08}, we visualize feature embeddings for both training samples and test samples.
Training samples in the first 5 class labels have low privacy risk scores, and training samples in the last 5 class labels have high privacy risk scores.
}\vspace{-2mm}
\label{fig:tsne_all}
\end{figure*}

\noindent \textbf{Privacy risk score with feature embeddings}
From the above experiment, we know that training samples from class labels with high generalization errors tend to have high privacy risk scores.
Next, we investigate this further by looking into the feature representations of different class labels learned by the target classifier.
We use the outputs of last \emph{hidden} layer of the target classifier as the feature embedding of the input sample.
We pick the top 5 class labels (30, 93, 97, 18, 98) with lowest average privacy risk scores (0.50, 0.52, 0.53, 0.54, 0.55) and at least 100 training samples, and the top 5 class labels (72, 49, 45, 51, 78) with highest average privacy risk scores (0.82, 0.83, 0.83, 0.85, 0.90) and at least 100 training samples.
We record feature embeddings for both training and test examples from these 10 class labels.
Finally, we adopt the t-Distributed Stochastic Neighbor Embedding (t-SNE) \cite{tsne_maaten_JMLR08}, a nonlinear dimensionality reduction technique, to visualize the feature embeddings.

Figure~\ref{fig:tsne_train} and Figure~\ref{fig:tsne_test} show the t-SNE plots of training samples and test samples, respectively.
The training samples are separated clearly based on class labels since the target classifier has the training accuracy close to $100\%$.
Test samples from class labels with low risk scores (classes 30, 93, 97, 18, 98) have quite similar feature embeddings as training samples and are still well separated.
On the other hand,
\textbf{test samples from class labels with high risk scores (classes 72, 49, 45, 51, 78) exhibit differences in feature representations compared to corresponding training samples.}
From Figure \ref{fig:tsne_all}, we also observe the heterogeneity of samples’ privacy risks, in the granularities of both individual samples (e.g., different samples in class 78) and class labels (e.g., class 30 versus class 78). This further emphasizes the importance of fine-grained privacy risk analysis. It also validates our attack design of using class-dependent thresholds in Section \ref{subsec:benchmark_attacks}.
Our observations are also important for future defense work. A good defense approach should make training data and validation data have similar feature embeddings and consider the heterogeneity of samples’ privacy risks.

\section{Conclusions}

In this paper, we first argue that measuring membership inference privacy risks with neural network based attacks is insufficient.
We propose to use a suite of metric-based attacks, including existing methods with our improved class-specific thresholds and a new proposed method based on modified prediction entropy, for benchmarking privacy risks of machine learning models.
We also make recommendations of comparing with early stopping when benchmarking a defense that introduces a tradeoff between model accuracy and privacy risks, and considering adaptive attackers with knowledge of the defense to rigorously evaluate the performance of defense approaches.  
With these benchmark attacks, we show that (1) the defense approach of adversarial regularization, proposed by Nasr et al. \cite{nasr_membership_defense_CCS18}, only reduces privacy risks to a limited degree and is no better than early stopping;
(2) the defense performance of MemGuard, proposed by Jia et al. \cite{jia2019memguard_ccs19}, is greatly degraded with adaptive attacks.

Next, we introduce a new metric called the privacy risk score for a fine-grained analysis of individual samples' privacy risks.
We show the effectiveness of the privacy risk score in estimating the true likelihood of an individual sample being in the training set and observe the heterogeneity of samples' privacy risk scores with experimental results.
Finally, we perform an in-depth investigation about the correlation between privacy risks and model properties, including sensitivity, generalization error, and feature embeddings.
We hope that our work convinces the research community about the importance of systematically and rigorously evaluating privacy risks of machine learning models.

\section*{Acknowledgements}
We are grateful to anonymous reviewers at USENIX Security for valuable feedback. 
We would also like to thank Google's TensorFlow Privacy team for integrating our methods.
This work was supported in part by the National Science Foundation under grants CNS-1553437 and CNS-1704105, the ARL's Army Artificial Intelligence Innovation Institute (A2I2), the Office of Naval Research Young Investigator Award, the Army Research Office Young Investigator Prize, Faculty research award from Facebook, Schmidt DataX award, and by Princeton E-ffiliates Award.

\bibliographystyle{plain}
\bibliography{references.bib}

\appendix

\section{Membership inference attacks against other datasets}\label{appendix:attacks_other_data}

Here, we perform membership inference attacks on two more image datasets: CH-MNIST and Car196.
The CH-MNIST dataset contains histology tiles from patients with colorectal cancer.\footnote{\url{https://www.kaggle.com/kmader/colorectal-histology-mnist}}
The dataset contains 64$\times$64 black-and-white images from 8 different classes of tissue, 5,000 samples in total. We use 2,000 data samples to train a convolution neural network. The model contains 2 convolution blocks with the number of output channels equal to 32 and 64.
The classifier achieves 99.0\% training accuracy and 71.7\% test accuracy.

The Car196 dataset contains colored images of 196 classes of cars.\footnote{\url{https://ai.stanford.edu/~jkrause/cars/car_dataset.html}}
The dataset is split into 8,144 training images and 8,041 testing images.
To train a model with good accuracy, we use a public ResNet50 \cite{he_ResNet_CVPR16} classifier pretrained on ImageNet \cite{deng2009imagenet} and fine-tune it on the Car196 training set.
The classifier achieves 99.3\% training accuracy and 87.5\% test accuracy.

\begin{table}[!ht]
\caption{membership inference attacks against image datasets}\vspace{-2mm}
\centering
\renewcommand\arraystretch{1.0}
\resizebox{\linewidth}{!}{
\begin{tabular}{cccccc}
\toprule[1.5pt]
\multirow{2}{*}{{dataset}} & {attack acc}  & {attack acc}  & {attack acc}  & {attack acc} & {attack acc}   \\
&  { (NN-based) } & { ($\mathcal{I}_{\text{corr}}$)} & { ($\mathcal{I}_{\text{conf}}$)} & { ($\mathcal{I}_{\text{entr}}$)} & { ($\mathcal{I}_{\mentropy}$)} \\
\midrule[0.75pt]

\multirow{2}{*}{CH-MNIST} &  \multirow{2}{*}{{70.5\%}} & \multirow{2}{*}{63.7\%} & \multirow{2}{*}{\textbf{72.6\%}} & \multirow{2}{*}{69.6\%} & \multirow{2}{*}{\textbf{72.6\%}}\\
   &  & & & & \\
\multirow{2}{*}{Car196} &  \multirow{2}{*}{{63.1\%}} & \multirow{2}{*}{55.9\%} & \multirow{2}{*}{\textbf{63.7\%}} & \multirow{2}{*}{62.9\%} & \multirow{2}{*}{\textbf{63.7\%}}\\
   &  & & & & \\
\bottomrule[1.5pt]
\end{tabular}
}
\vspace{-0.5em}
\label{tab:attacks_more_data}
\end{table}

Besides our benchmark attacks, we follow Nasr et al. \cite{nasr_whitebox_privacy_SP19} to perform NN-based attacks.
We present attack results in Table~\ref{tab:attacks_more_data}. 
We can see that the best attack accuracy of our benchmark attacks is $2.1\%$ and $0.6\%$ larger than NN-based attacks.

\section{Privacy risk score with different training/test selection probabilities}\label{appendix:different_priors}

\begin{figure}[!ht]
\centering
	\begin{subfigure}[t]{0.47\linewidth}
		\raggedleft
		\includegraphics[width=\linewidth]{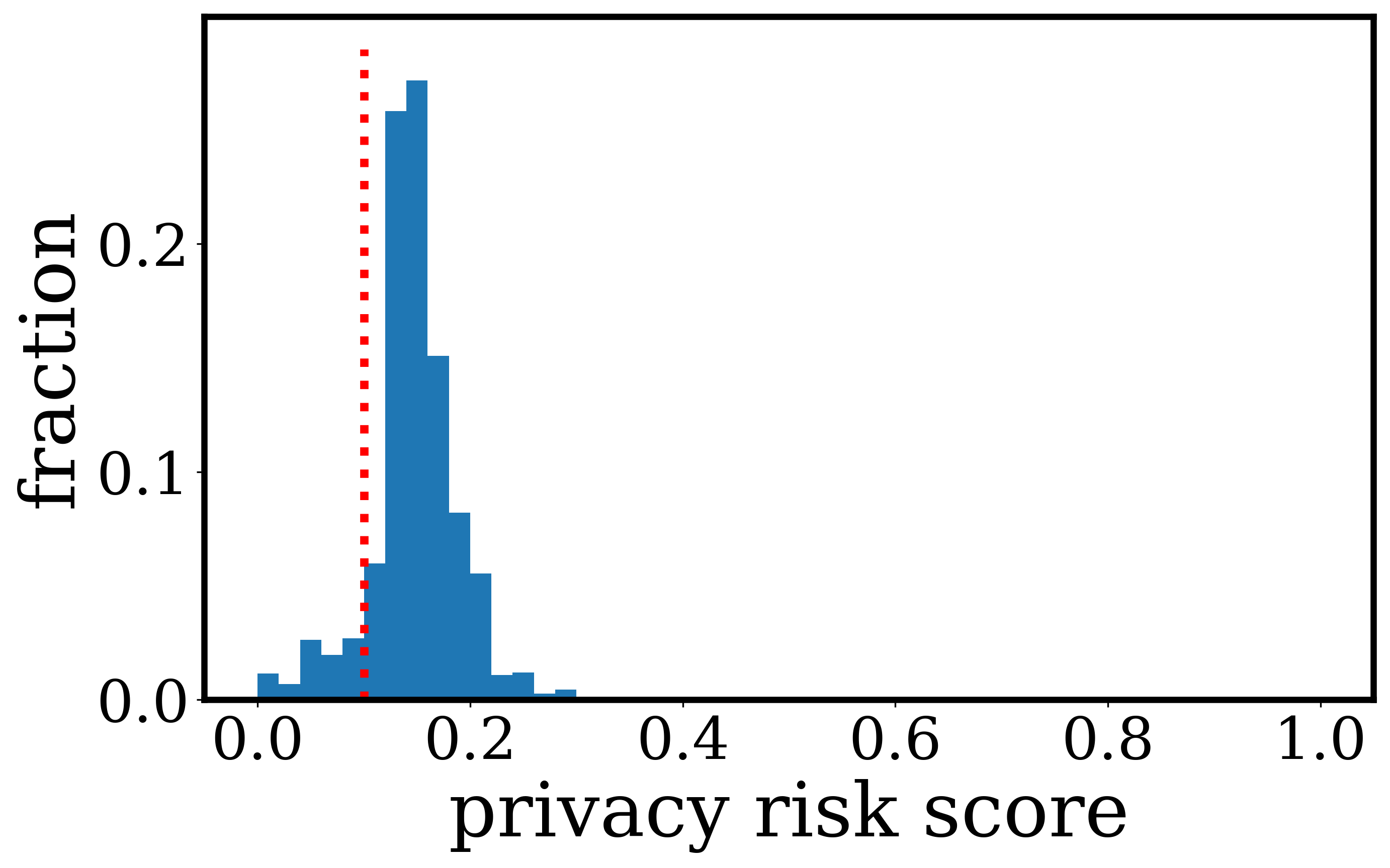}
		\caption{$P(\bfz \in D_{\text{tr}}) = 0.1$}
	\end{subfigure}\hfill
	\begin{subfigure}[t]{0.47\linewidth}
		\raggedright
		\includegraphics[width=\linewidth]{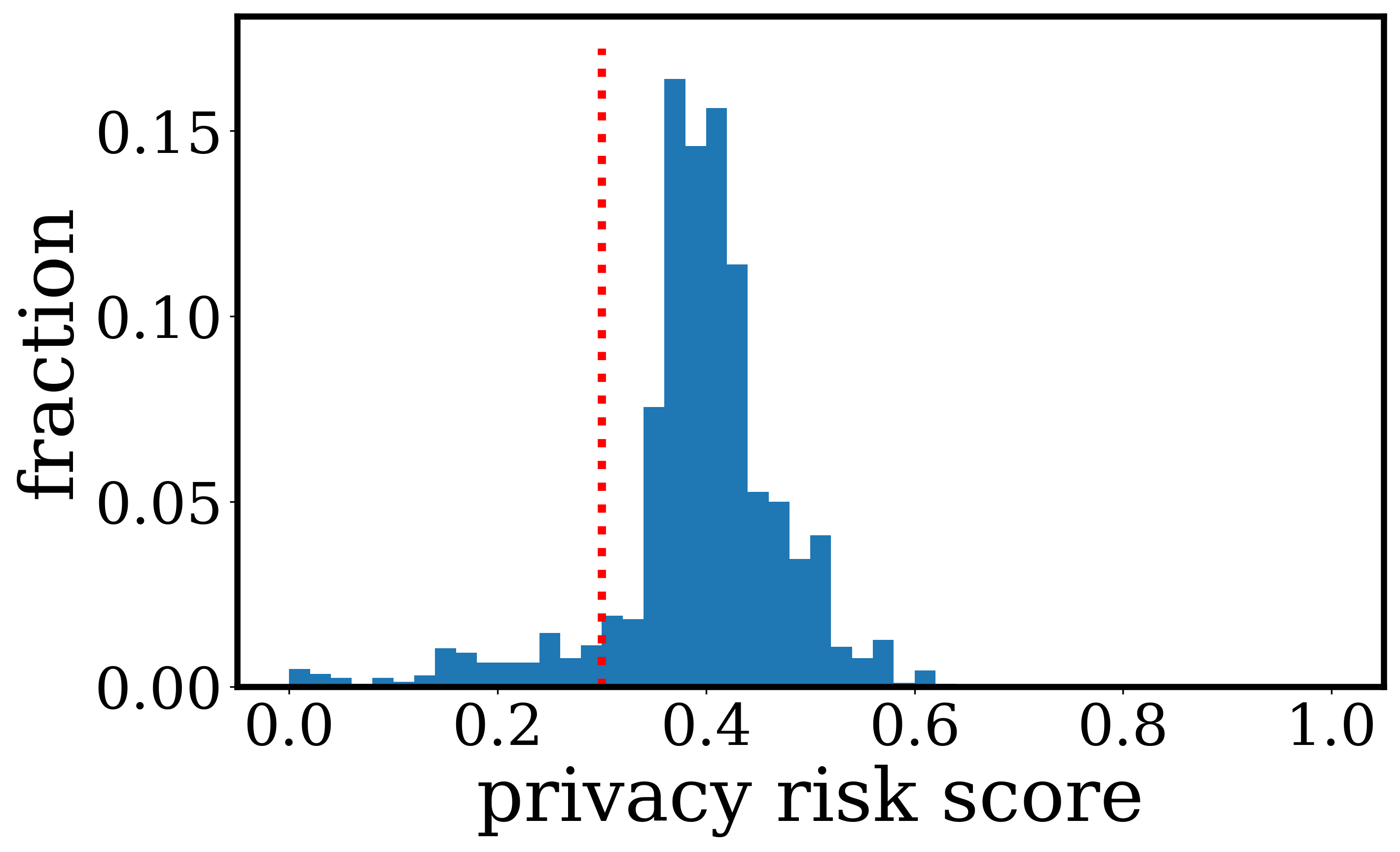}
		\caption{$P(\bfz \in D_{\text{tr}}) = 0.3$}
	\end{subfigure}\hfill
	\begin{subfigure}[t]{0.47\linewidth}
		\raggedleft
		\includegraphics[width=\linewidth]{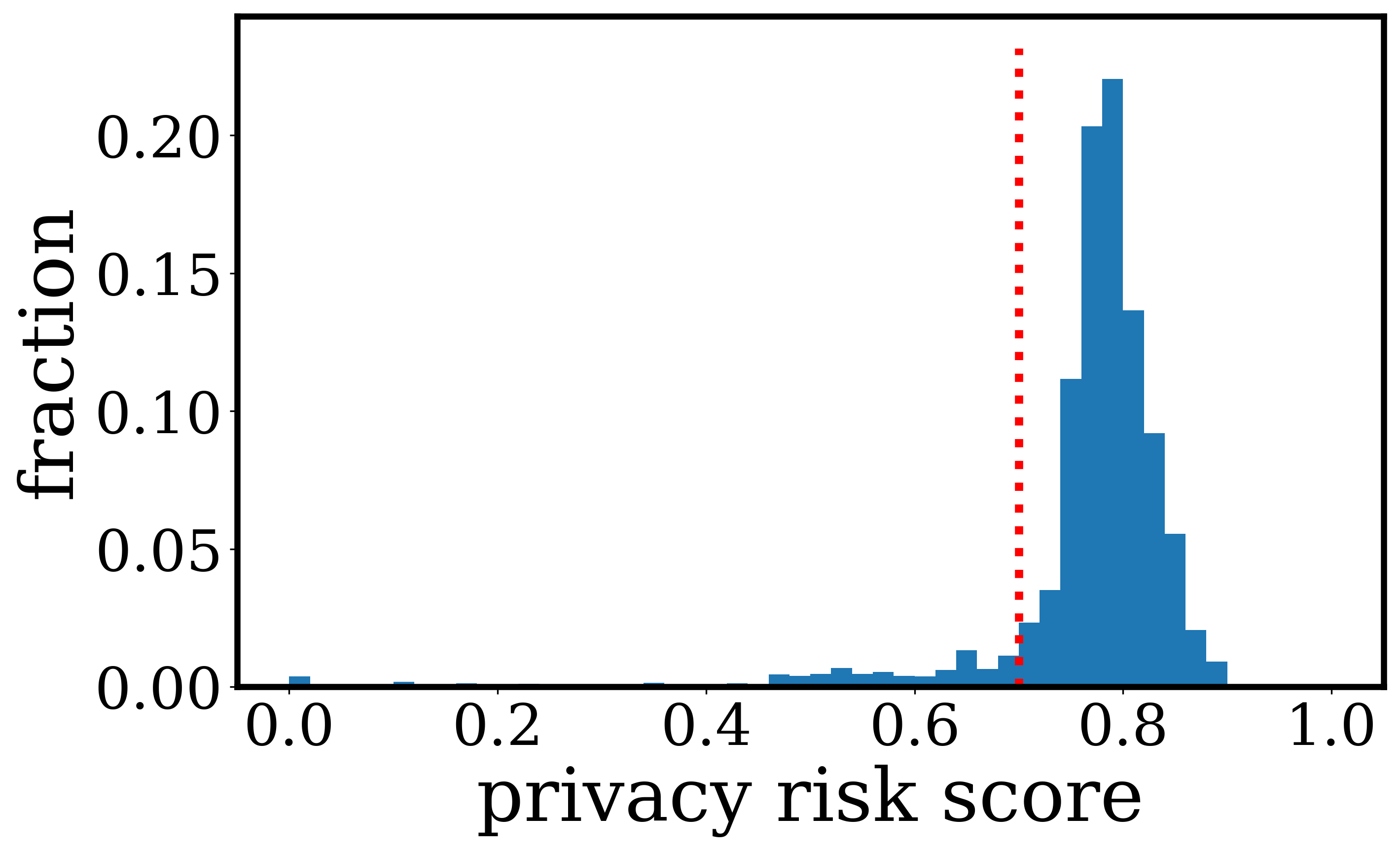}
		\caption{$P(\bfz \in D_{\text{tr}}) = 0.7$}
	\end{subfigure}\hfill
	\begin{subfigure}[t]{0.47\linewidth}
		\raggedright
		\includegraphics[width=\linewidth]{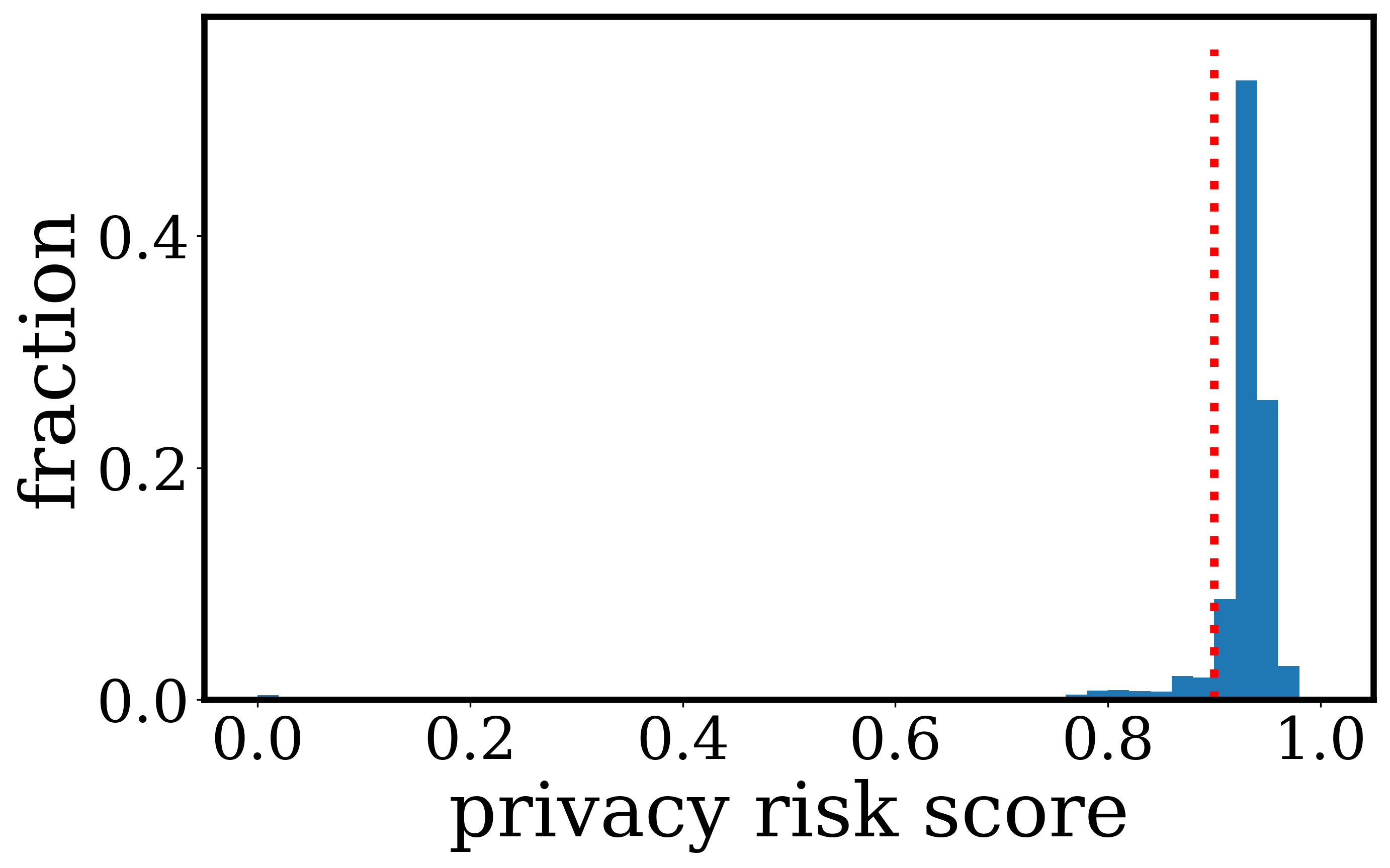}
		\caption{$P(\bfz \in D_{\text{tr}}) = 0.9$}
	\end{subfigure}\hfill
\vspace{-0.5em}
\caption{
For the undefended Purchase100 classifier, we present the distribution of training data' privacy risk scores with varied training set prior probability $P(\bfz \in D_{\text{tr}})$. For each figure, we also plot the baseline of prior probability.
}\vspace{-3mm}
\label{fig:risk_score_varied_prior}
\end{figure}

Here, we provide the privacy risk score results on undefended Purchase100 classifier when the sample is chosen from training or test set with different probabilities.
The computation of privacy risk score ($r(\bfz)$) is same as Section \ref{subsec:risk_score}, except we use Equation \eqref{eq:risk_member_2} by also considering prior distributions $P(\bfz \in D_{\text{tr}})$ and $P(\bfz \in D_{\text{te}})=1-P(\bfz \in D_{\text{tr}})$.
We present the results in Figure \ref{fig:risk_score_varied_prior} with different values of $P(\bfz \in D_{\text{tr}})$, where the red dotted line represents the baseline of random guessing.
We can see that in all cases, most training samples have privacy risk scores larger than the prior training probability. 
We further compute a \emph{distance value} between the prior distribution and the privacy risk score (posterior) distribution  as $\frac{1}{|D_{\text{tr}}|}\sum_{\bfz \in D_{\text{tr}}} (r(\bfz)-P(\bfz \in D_{\text{tr}}))$ to represent the privacy leakage. The distance values are 0.05, 0.09, 0.07, 0.02 when $P(\bfz \in D_{\text{tr}}) = 0.1, 0.3, 0.7, 0.9$, respectively. As a comparison, the distance value is 0.1 when $P(\bfz \in D_{\text{tr}}) = 0.5$.
As $P(\bfz \in D_{\text{tr}})$ is closer to 0.5, the uncertainty of membership inference is larger, which in turns leads to a larger distance value.

\section{Validation of privacy risk score on Texas100 classifiers}
\label{appendix:texas_score_valid}

\begin{figure}[!ht]
\centering
	\begin{subfigure}[t]{0.47\linewidth}
		\raggedleft
		\includegraphics[width=\linewidth]{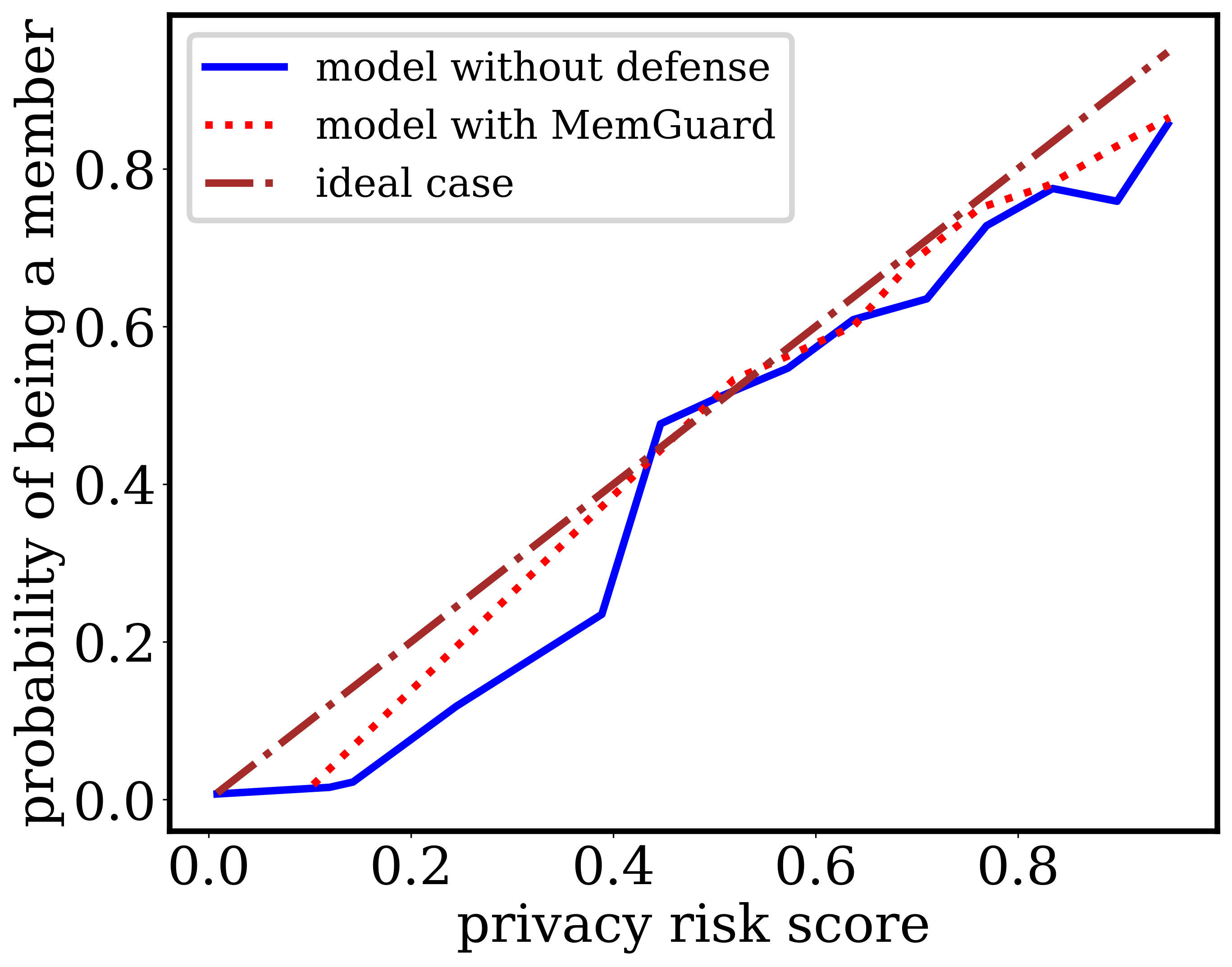}
	\end{subfigure}\hfill
	\begin{subfigure}[t]{0.47\linewidth}
		\raggedright
		\includegraphics[width=\linewidth]{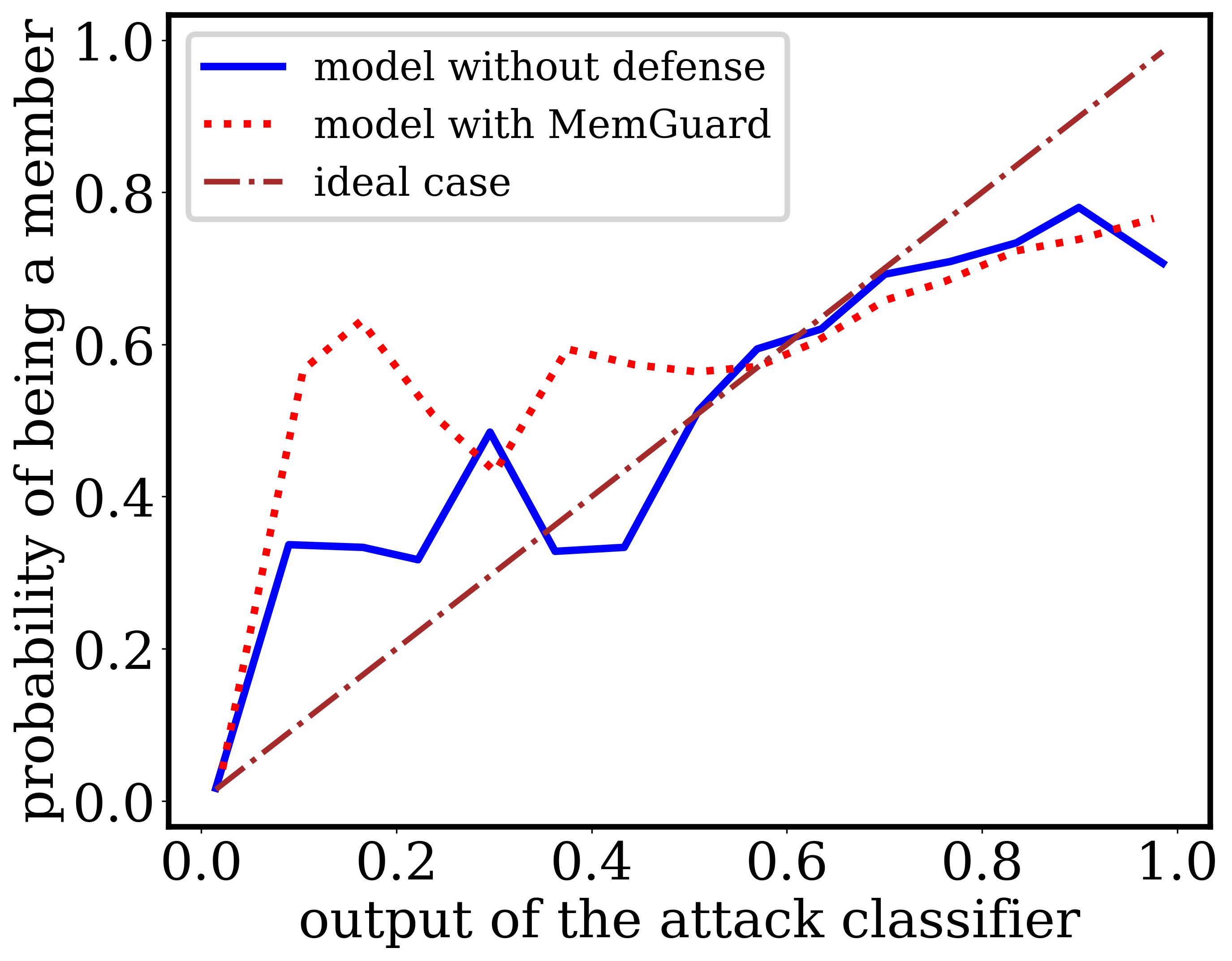}
	\end{subfigure}\hfill
\vspace{-0.5em}
\caption{
For Texas100 classifiers, estimate the real probability of being a member by using our proposed privacy risk score (left) and using the output of the NN attack classifier (right). The root-mean-square errors (RMSE) values of our privacy risk score are 0.08 and 0.05, while the RMSE values of NN attack classifier's output are 0.13 and 0.21.
}\vspace{-3mm}
\label{fig:risk_score_valid_texas}
\end{figure}

We validate the effectiveness of privacy risk score on the undefended Texas100 classifier and its defended version with MemGuard \cite{jia2019memguard_ccs19} in Figure~\ref{fig:risk_score_valid_texas}.
Compared with the output of NN attacks, our proposed privacy risk score is more meaningful for indicating the real probability of being a member.
The RMSE values with privacy risk score are 0.08 and 0.05, while the RMSE values with NN classifier outputs are 0.13 and 0.21, for the undefended and defended Texas100 classifiers.

\section{Validation of privacy risk scores on different model architectures}\label{appendix:validation_varied_architecture}

\begin{figure}[!ht]
	\centering

	\begin{subfigure}[t]{0.47\linewidth}
		\raggedleft
		\includegraphics[width=\linewidth]{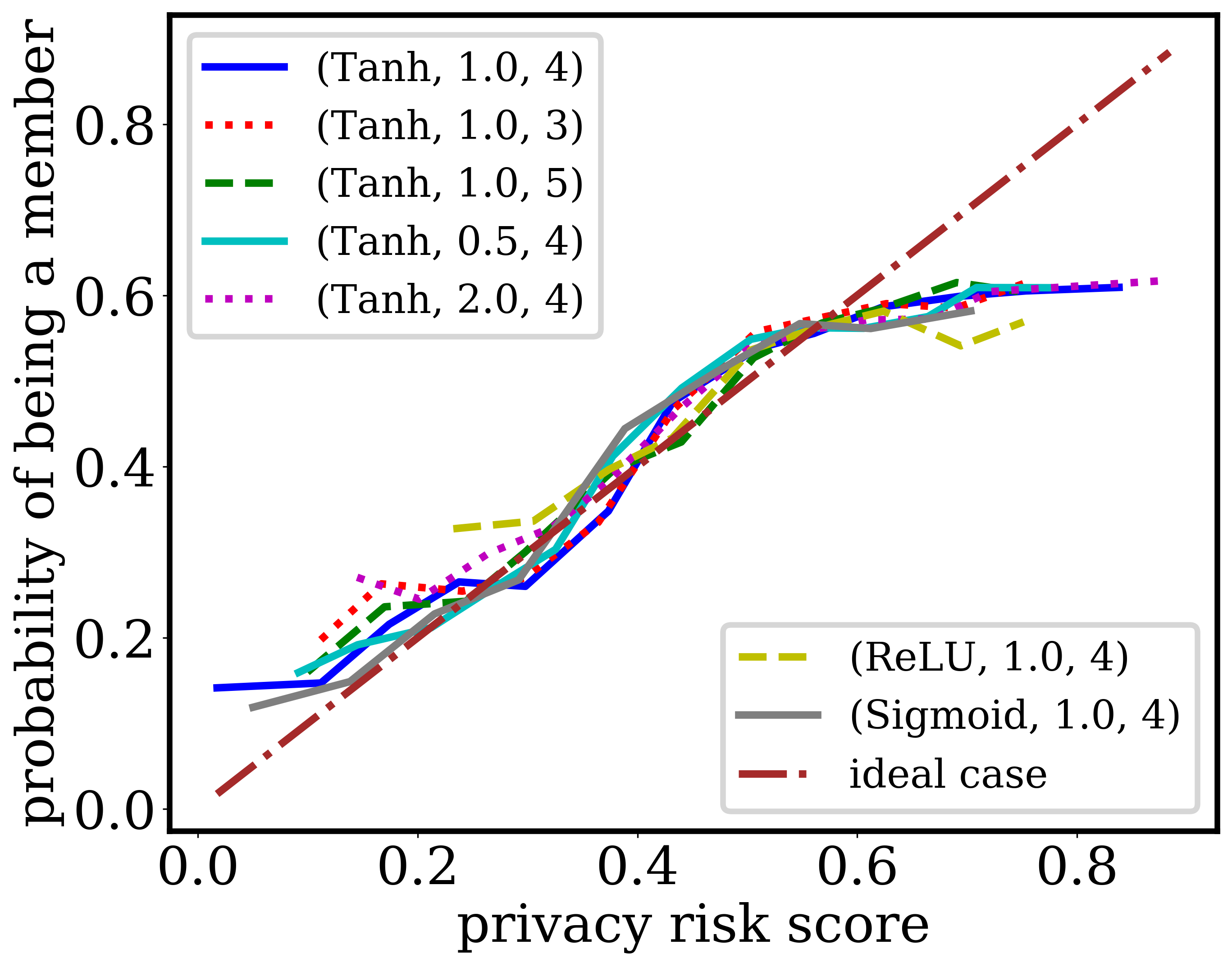}
	\end{subfigure}\hfill
	\begin{subfigure}[t]{0.47\linewidth}
		\raggedright
		\includegraphics[width=\linewidth]{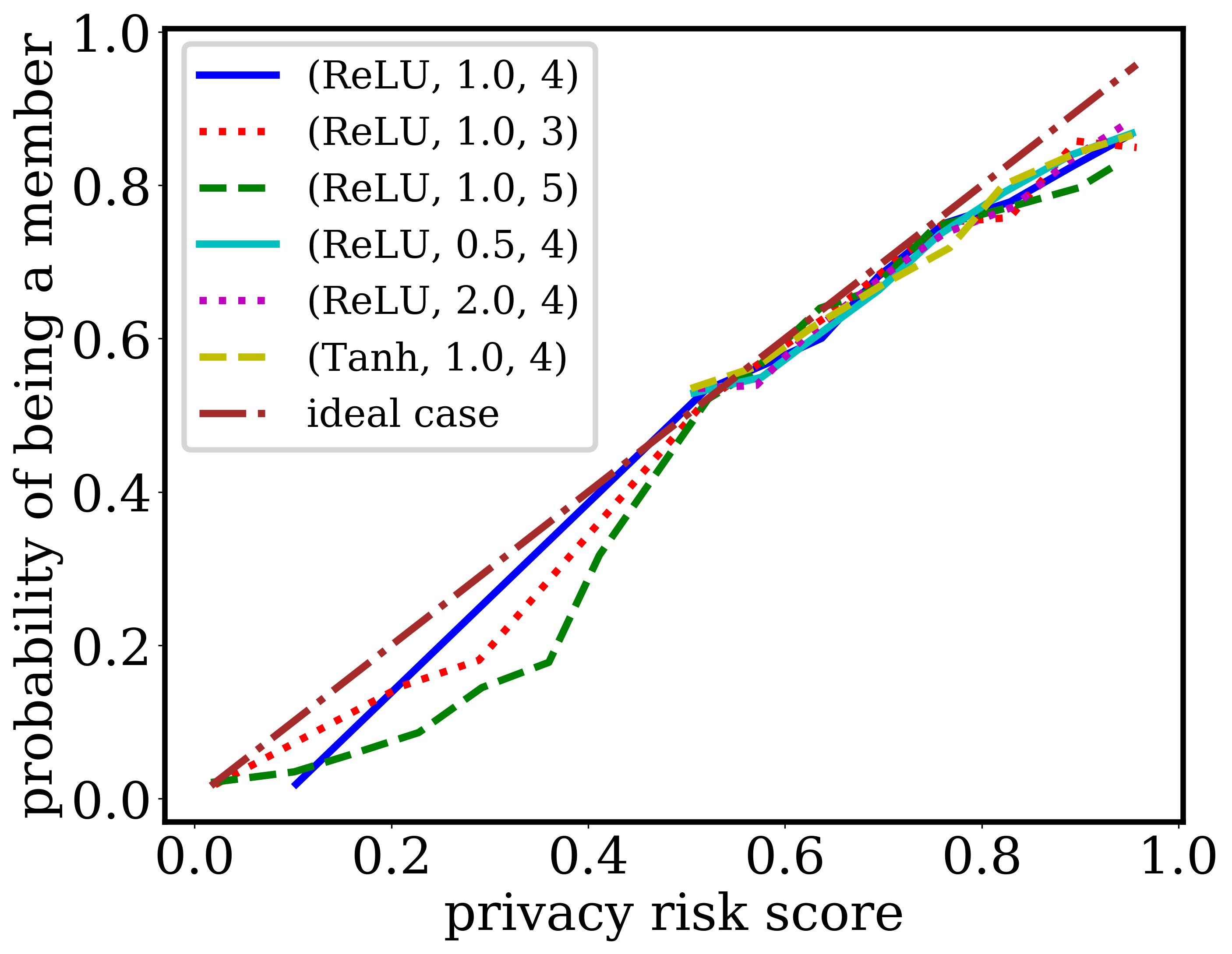}
	\end{subfigure}\hfill
	\vspace{-0.5em}
	\caption{Validation of privacy risk score with varied model architectures on defended Purchase100 classifiers \cite{nasr_membership_defense_CCS18} (left) and defended Texas100 classifiers \cite{jia2019memguard_ccs19} (right). 
	The legend is expressed as (activation function, width, depth). The RSME values between privacy risk score (x-axis) and probability of being a member (y-axis) for all lines are smaller than 0.10.}
	\label{fig:more_validation_appendix}\vspace{-3mm}
\end{figure}

We provide more validation results on Purchase100 classifiers defended by adversarial regularization \cite{nasr_membership_defense_CCS18} and Texas100 classifiers defended by MemGuard \cite{jia2019memguard_ccs19} in Figure \ref{fig:more_validation_appendix}.
We can see that for all lines, the privacy risk score is close to the probability of being a member.

\end{document}